\documentclass[sigconf]{acmart}

\renewcommand\footnotetextcopyrightpermission[1]{} 
\AtBeginDocument{%
  }

\setcopyright{acmcopyright}
\copyrightyear{2023}
\acmYear{2023}
\acmDOI{XXXXXXX.XXXXXXX}

\acmConference[Conference acronym 'XX]{Make sure to enter the correct
  conference title from your rights confirmation email}{June 03--05,
  2023}{Woodstock, NY}
\acmPrice{15.00}
\acmISBN{978-1-4503-XXXX-X/18/06}

\usepackage{algorithm}  
\usepackage[noend]{algpseudocode}
\usepackage{algorithmicx,algorithm}
\usepackage{makecell}
\usepackage{multirow} 
\usepackage{listings}
\usepackage{textcomp}
\usepackage{color}
\usepackage{colortbl}
\definecolor{codegray}{rgb}{0.5,0.5,0.5}
\definecolor{codepurple}{HTML}{C42043}
\definecolor{backcolour}{HTML}{F2F2F2}
\definecolor{bookColor}{cmyk}{0,0,0,0.90} 
\color{bookColor}
\usepackage{pifont}
\usepackage{listings}
\AtBeginDocument{\DeclareCaptionSubType{lstlisting}}
\usepackage{caption,subcaption}

    \lstdefinestyle{mystyle}{
        backgroundcolor=\color{backcolour},   
        commentstyle=\color{codepurple},
        numberstyle=\footnotesize\color{codegray},
        basicstyle=\footnotesize,
        breakatwhitespace=false,         
        breaklines=true,                 
        captionpos=b,                    
        keepspaces=true,                 
        numbersep=-10pt,                  
        showspaces=false,                
        showstringspaces=false,
        showtabs=false,      
    }
\begin{document}

\title{Detecting Logic Bugs of Join Optimizations in DBMS}


\author{Xiu Tang}
\affiliation{
  \institution{Zhejiang University}
  \country{}
  }
\email{tangxiu@zju.edu.cn}

\author{Sai Wu}
\authornote{Sai Wu is the corresponding author.}
\affiliation{
  \institution{Zhejiang University}
  \country{}
  }
\email{wusai@zju.edu.cn}

\author{Dongxiang Zhang}
\affiliation{
  \institution{Zhejiang University}
  \country{}
  }
\email{zhangdongxiang@zju.edu.cn}

\author{Feifei Li}
\affiliation{
  \institution{Alibaba Group}
  \country{}
  }
\email{lifeifei@alibaba-inc.com}

\author{Gang Chen}
\affiliation{
  \institution{Zhejiang University}
  \country{}
  }
\email{cg@zju.edu.cn}

\renewcommand{\shortauthors}{Xiu Tang, et al.}

\begin{abstract}
  Generation-based testing techniques have shown their effectiveness in detecting logic bugs of DBMS, which are often caused by improper implementation of query optimizers. Nonetheless, existing generation-based debug tools are limited to single-table queries and there is a substantial research gap regarding multi-table queries with join operators.
  In this paper, we propose TQS, a novel testing framework targeted at detecting logic bugs derived by queries involving multi-table joins. Given a target DBMS, TQS achieves the goal with two key components: Data-guided Schema and Query Generation (DSG) and Knowledge-guided Query Space Exploration (KQE). DSG addresses the key challenge of multi-table query debugging: how to generate ground-truth (query, result) pairs for verification.  
  It adopts the database normalization technique to generate a testing schema and maintains a bitmap index for result tracking. To improve debug efficiency, DSG also artificially inserts some noises into the generated data. 
  To avoid repetitive query space search, KQE forms the problem as isomorphic graph set discovery and
  combines the graph embedding and weighted random walk for query generation. We evaluated TQS on four popular DBMSs: 
  MySQL, MariaDB, TiDB and 
  the gray release of an industry-leading cloud-native database, anonymized as X-DB.
  Experimental results show that TQS is effective in finding logic bugs of join optimization in database management systems. It successfully detected 
  115
  bugs within 24 hours,
  including 
  31
  bugs in MySQL, 
  30
  in MariaDB, 31 in TiDB, and 23 in X-DB respectively. 
\end{abstract}

\ccsdesc[300]{Information systems~Data management systems}
\ccsdesc[300]{Security and privacy~Database and storage security}
\ccsdesc[300]{Information systems~Structured Query Language}

\keywords{Database, logic bug, join optimization.}

\maketitle

\section{Introduction}
 In the past decades, we have witnessed the evolution of modern DBMS (Database Management Systems) to support various new architectures such as cloud platforms and HTAP~\cite{DBLP:conf/sigmod/CaoZYLWHCCLFWWS21, huang2020tidb}, which require increasingly sophisticated optimizations for query evaluation. Query optimizer is considered as one of the most complex and important components in DBMS. It parses the input SQL queries and generates an efficient execution plan with the assistance of built-in cost models. The implementation errors in a query optimizer can result in bugs, including crashes and logic bugs. Crashes are easier to detect as the system will halt immediately. Whereas, logic bugs are prone to be ignored, because they simply cause the DBMS to return incorrect result sets that are hard to detect. In this paper, we focus on detecting these silent bugs.

\begin{figure}
    \centering
    \subcaptionbox{MySQL's incorrect hash join execution.}{
        \centering
        \includegraphics[scale=0.12]{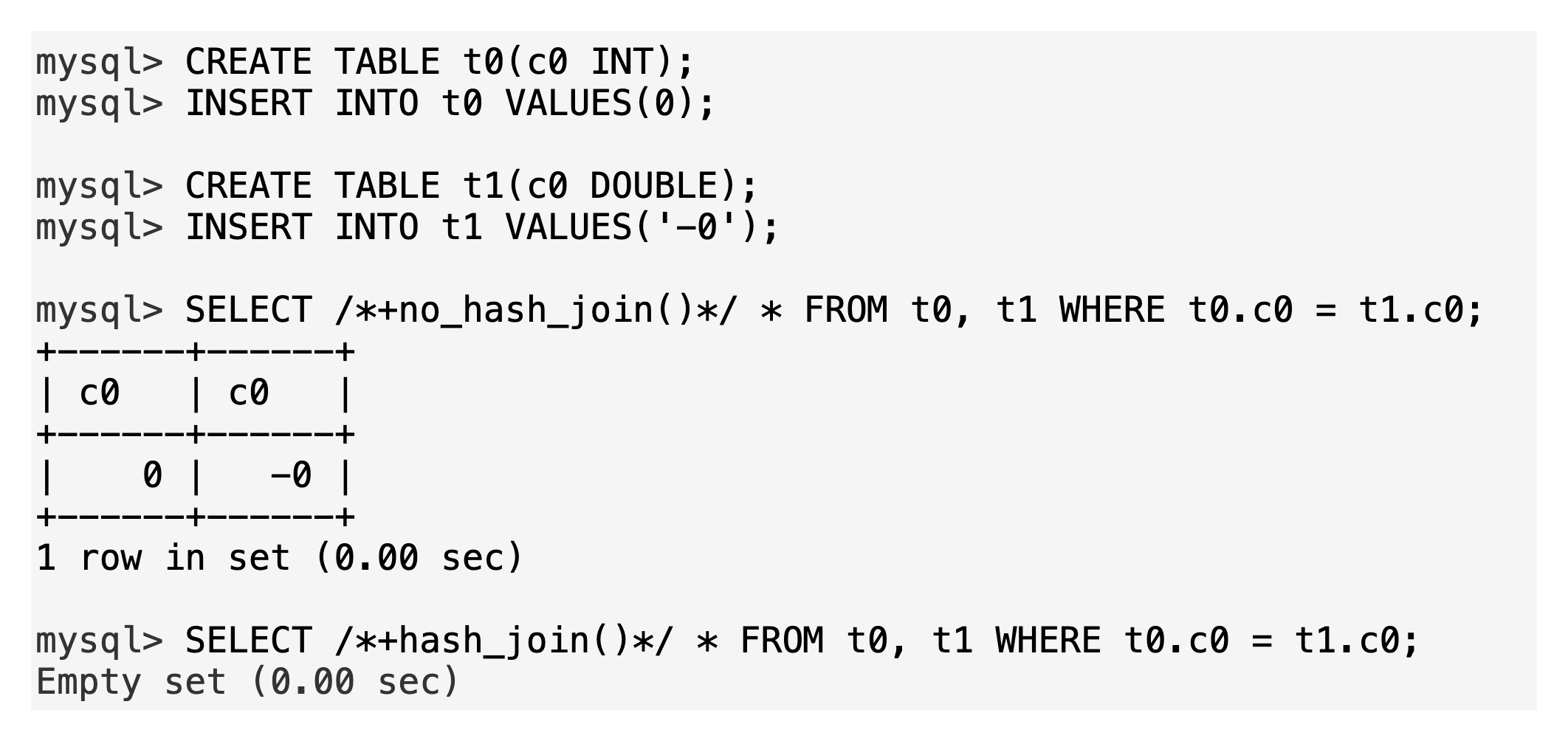}
    }
     \subcaptionbox{MySQL's incorrect semi-join execution.}{
        \centering
        \includegraphics[scale=0.12]{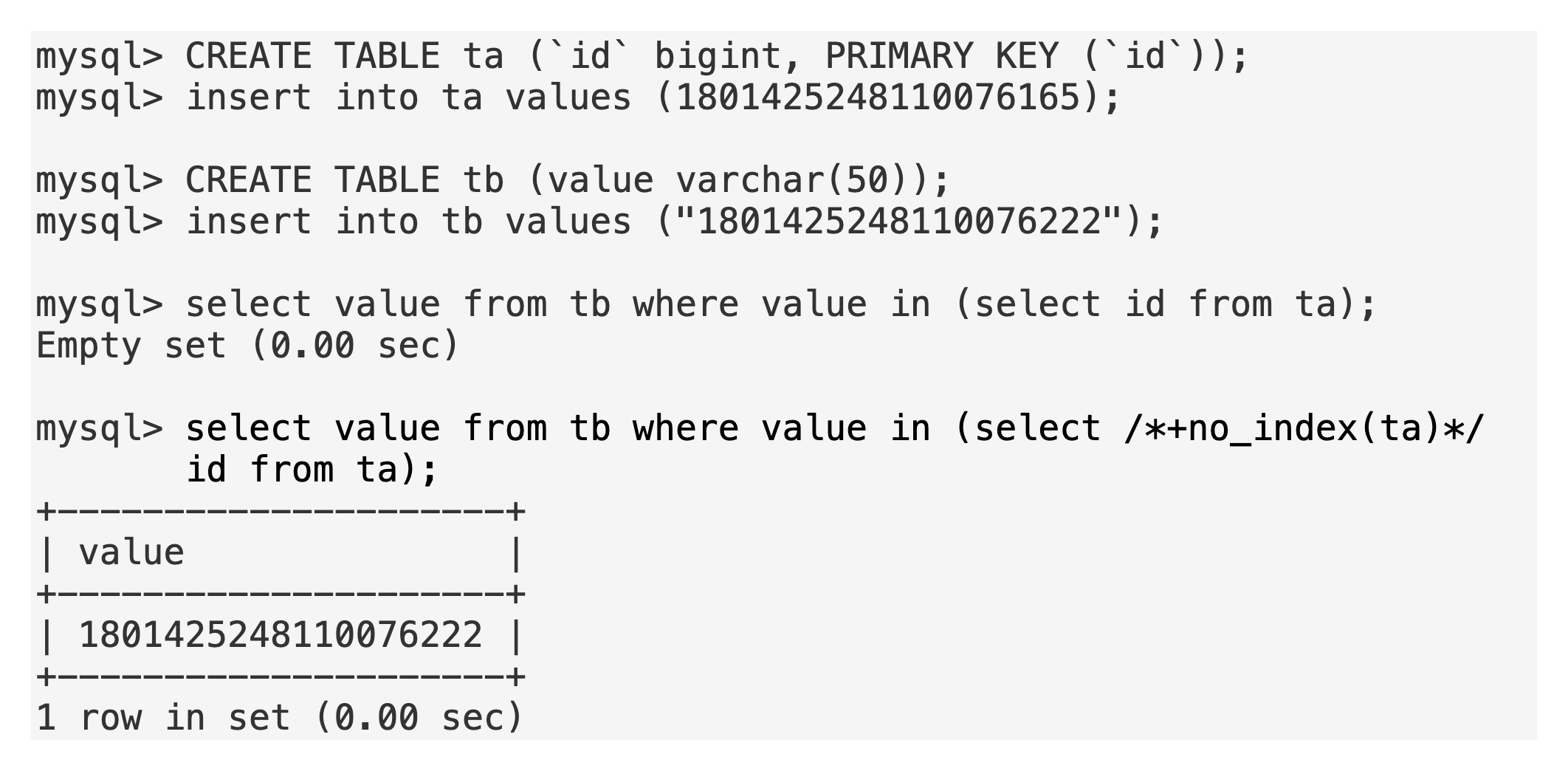}
    }
    \vspace{-0.3cm}
    \caption{Logic bug cases of join optimizations in MySQL.}
    \vspace{-0.4cm}
    \label{fig.intro}
  \end{figure}
  
  \begin{figure*}
  \centering
  \includegraphics[scale=1]{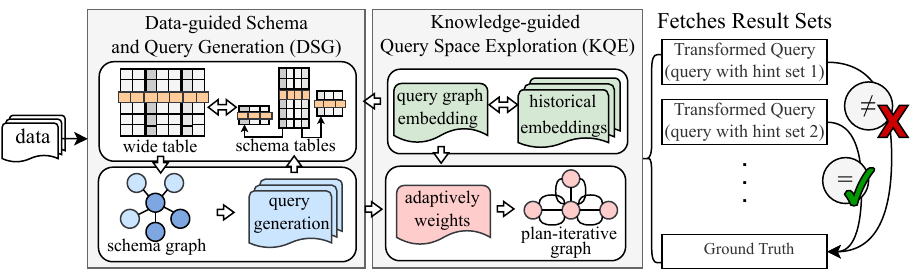}
   \vspace{-0.2cm} 
  \caption{Overview of TQS. TQS designs DSG (Data-guided Schema and query Generation) and
  KQE (Knowledge-guided Query space Exploration)
  to detect the logic bugs of join optimizations in DBMS.}
\vspace*{-0.2cm} 
  \label{fig.overview}
\end{figure*}

Pivoted Query Synthesis (PQS) has recently emerged as a promising way to detect logic bugs in DBMS~\cite{rigger2020testing}. Its core idea is to select a pivot row from a table and generate queries that fetch this row as the result. A logic bug is detected if the pivot row is not returned in any synthesized query. PQS is mainly designed to support selection queries in a single table and $90\%$ of its reported bugs are for queries involving only one table. There still exists substantial research gap regarding multi-table queries with different join algorithms and join structures, which are more error-prone than single-table queries.

In Figure \ref{fig.intro}, we illustrate two logic bugs of MySQL for join queries. 
These two bugs can be detected by our proposed tool in this paper.
Figure \ref{fig.intro}(a) demonstrates a logic bug of hash join in MySQL 8.0.18.
In this example, the first query returns the correct result set because it is executed with the block nested loop join.
However, when the second query is issued with an inner hash join,  an incorrect empty result set is returned. This is because the underlying hash join algorithm asserts that ``$0$'' and ``$-0$'' are not equal.
In Figure \ref{fig.intro}(b), the logic bug is caused by semi-join processing in MySQL's newest version (8.0.28).
In the first query, the nested loop inner join casts the data type $varchar$ to $bigint$ and produces a correct result set.
But when the second query is executed with hash semi-join, the data type $varchar$ is converted to $double$, resulting in data accuracy loss and incorrect equivalence comparison.

Adopting query synthesis for logic bug detection in multi-table join queries is much more difficult than that in single-table selection queries, due to two unique challenges:
\begin{itemize}
\item 
{\bf Result Verification}: Previous approaches adopt the
differential testing strategy to verify the correctness of
query results. The idea is to process a query using different physical plans. If plans return inconsistent result sets, a possible logic bug is detected. However, the drawback of differential testing is two-fold. First, some logic bugs affect multiple physical plans and make them all generate the same incorrect result. Second, when inconsistent result sets are observed, we need manually check which plan generates the correct one, incurring high overheads. A possible solution for the above problem is to obtain the ground-truth results for an arbitrary testing query, which is not supported by existing tools. 
\item  {\bf Search Space}: The number of join queries that can be generated from a given database schema is exponential to the number of tables and columns. Since we are unable to enumerate all possible queries for verification, there requires an effective query space exploration mechanism that allows us to detect logic bugs as efficiently as possible.
\end{itemize}

In this paper, we propose \textbf{T}ransformed \textbf{Q}uery \textbf{S}ynthesis (TQS) as a remedy. It is a novel, general, and cost-effective tool to detect logic bugs of join optimizations in DBMS. To address the first challenge above, we propose the DSG (Data-guided Schema and query Generation) approach. Given a dataset denoted as one wide table, DSG splits the dataset into multiple tables based on detected normal forms. 
To speed up bug discovery, DSG also inserts some artificial noise data into the generated database.
We first convert the database schema into a graph whose nodes are the tables/columns and edges are the relationships between the nodes.
DSG adopts random walking on schema graph to select tables for queries, and uses those tables to generate join expressions. 
For a specific join query spanning over multiple tables, we can easily identify its ground-truth results from the wide table. In this way, DSG can effectively generate (query, result) pairs for database verification.

For the second challenge, we design the KQE 
(Knowledge-guided Query space Exploration) approach. 
We first extend the schema graph to a plan-iterative graph, which represents the entire query space.
Each join query is then represented as a sub-graph. 
KQE adopts an embedding-based graph index to score the generated query graphs by searching whether there are structurally similar query graphs in already-explored space. The coverage score guides the random walk generator to explore the unknown query space as much as possible.

To demonstrate the generality and effectiveness of our approach, we evaluated TQS on four popular DBMSs: MySQL \cite{mysql}, MariaDB \cite{mariadb}, TiDB \cite{huang2020tidb}
and an industry-leading cloud native database (anonymized as X-DB).
After $24$ hours of running,
TQS successfully found 
$115$ 
bugs,
including 
$31$
bugs in MySQL, 
$30$
in MariaDB, $31$ in TiDB, and $23$ in X-DB. 
Through root cause analysis, there are 
$7$
types of bugs in MySQL, 
$5$ 
in MariaDB, $5$ in TiDB, and $3$ in X-DB respectively. 
All the detected bugs have been submitted to the respective community and in the future, we will open-source TQS as a database debugging tool.

\begin{table}[b]
  \centering
  \vspace{-0.3cm}
    \caption{Summary of notations.}
    \vspace{-0.2cm}
    \label{notation}
    \begin{tabular}{|c|l|}
    \hline
     Notation & Definition  \\
    \hline
        $d$ & input dataset \\ \hline
        $T_w$ & wide table which has multiple columns \\ \hline
        $S$ & generated database schema\\ \hline
        $G_s$ & database schema graph \\ \hline
        $G$   & plan-iterative graph \\ \hline
        $G_q$ & query graph \\ \hline
        $q_i$ & generated SQL query $q_i$\\ \hline
        $trans\_q$ & transformed SQL query with hints for query $q$ \\ \hline
        $S_{q}$ & query result set for a query $q$ \\\hline 
        $GT_{q}$ & ground truth of query $q$'s query result set\\ \hline
    \end{tabular}
    \end{table}

\section{Overview}
In this section, we formulate the problem definition and present an overview of our proposed solution. 

\subsection{Problem Definition}

There are two categories for database bugs: crash and logic bugs.
Crash bugs are raised either by the operating system, or by the process of DBMS. They cause DBMS to be forcefully killed or halted, due to limited resources (e.g., out of memory) or access to an invalid memory address, etc.
Crash bugs are easy to be noticed. In contrast, the logic bugs are much harder to be noticed, 
because database still runs normally and query processing returns seemingly correct results (and indeed they return correct results in most cases, but may fetch incorrect result sets in corner cases).
These silent bugs (acting as hidden bombs) are more dangerous, since they are hard to detect and may affect the correctness of applications.

In this paper, we focus on detecting logic bugs introduced by the query optimizer 
for multi-table join queries. 
Specifically, we refer to those bugs as join optimization bugs.
Using the notations listed in Table \ref{notation}, the {\em join optimization bug detection} is formally defined as:

\begin{definition}
For each query $q_i$ in the query workload $Q$, we let the query optimizer execute 
joins of
$q_i$ with multiple physical plans,
and verify its result sets $S_{q_i}$ with its ground truth $GT_{q_i}$.
If $S_{q_i} \neq GT_{q_i}$, we find a join optimization bug.
\end{definition}

\subsection{Scheme Overview}

Figure \ref{fig.overview} shows an overview for the architecture of TQS.
Given a benchmark dataset and target DBMS, TQS searches for possible logic bugs of the DBMS by issuing queries against the dataset.
TQS achieves its goal with two key components: {\em Data-guided Schema and query Generation (DSG)} and {\em Knowledge-guided Query space Exploration (KQE)}.

DSG considers the input dataset as a wide table, and besides the original tuples, DSG deliberately synthesizes some tuples with error-prone values (e.g., null values or very long strings).
To target at join queries, DSG crafts a new schema for the wide table by splitting the wide table into multiple tables with normal form guarantees based on functional dependency. 
DSG models the database schema as a graph and generates logic/conceptual queries via random walk on the schema graph.
DSG materializes a logic query into physical plans, and transforms the query with different hints to enable the DBMS to execute multiple different physical plans for bug searching.
The ground-truth results of a join are identified by mapping the join graph back to the wide table.

After the schema is setup and data are split, KQE extends the schema graph to a plan-iterative graph. 
Each query is represented as a sub-graph.
KQE builds an embedding-based graph index for the embeddings of query graph in history (i.e., in already explored query space). 
The intuition of KQE is to assure a newly generated query graph is as far away from its nearest neighbors in history as possible, namely, to explore new query graphs, instead of repeating existing ones.
KQE achieves its effectiveness by scoring the generated
query graphs based on their structural similarity (to query graphs in history) and applying an adaptive random walk method for generation.

We summarize the core idea of TQS in Algorithm \ref{alg:all}, where we show the two main components:
DSG (line 2, line 10 and line 12), KQE (line 4, line 8 and line 9), respectively.

Given a dataset $d$ and a wide table $T_w$ sampled from $d$,  DSG splits the single wide table $T_w$ 
into multiple tables that form the database schema $S$ with normal form guarantees (line 2).
Schema $S$ can be considered as a graph $G_s$, where tables and columns denote vertices and edges represent the relationships between them.
DSG applies random walk on $G_s$ to generate the join expressions of queries (line 10). In fact, a join query can be projected as a sub-graph of $G_s$.
By mapping the sub-graph back to the wide table $T_w$, DSG can easily retrieve the ground-truth results for the query (line 12).

\renewcommand{\algorithmicrequire}{\textbf{Input:}}
\renewcommand{\algorithmicensure}{\textbf{Output:}}
\renewcommand{\Return}{\textbf{Return:}}
\newcommand{\Break}{\State \textbf{break} }
\begin{algorithm}[t]  
\caption{TQS ($d$, $H$, $\gamma$, $l$)}  
\label{alg:all}
\begin{algorithmic}[1]
\Require dataset $d$, hint set $H$, walks per vertex $\gamma $, maximum walk length $l$.
\Ensure returns bug logs $bugs$, graph index {\em GI}.
\State Initialize weights of edges $\pi=\{1\}$, $bugs, GI$
\State $S = DBGen(d)$
\State $G_s(V, E) = Schema2Graph(S)$
\State $G(V, E, \pi) = PlanIterative(G_s)$
\For{ $i=0$ to $\gamma$}
   \State $O = Shuffle(V)$
   \For{ each $v_i \in O$}
      \State $W_{v_{i}} = AdaptiveRandomWalk(G, v_i, l, GI)$
      \State ${GI} = {GI} \cup Embedding(W_{v_{i}})$
      \State $q_i = QueryGen(W_{v_i})$
      \State $trans\_q_i = HintGen(q_i, H, G_t)$
      \State $GT_{q_i} = getGT(q_i)$
      \State $S_{trans\_q_i} = ResultSet(trans\_q_i)$
      \If{$ S_{trans\_q_i} \neq GT_{q_i}$}
        \State $bugs = bugs \cup trans\_q_i $
      \EndIf
   \EndFor
\EndFor
\State \Return $\ bugs$, $GI$ 
\end{algorithmic}  
\end{algorithm}

KQE extends the schema graph to a plan-iterative graph (line 4). 
To avoid tests on similar paths, KQE builds an embedding-based graph index {\em GI} to index embeddings of the existing query graphs (line 9).
KQE updates the edge weights $\pi$ of the plan-iterative graph $G$ according to how much the current query graph is structurally similar to existing query graphs (line 8).
KQE scores the next possible paths, which guides the random walk generator to favor exploring unknown query space. 

For a query $q_i$, TQS transforms the query with hint sets $trans\_q_i$ to execute multiply different physical query plans (line 11).
Finally, the result set of query $trans\_q_i$ is compared with the ground-truth $GT_{q_i}$ (line 14).
If they are not consistent, a join optimization bug is detected (lines 15).

TQS currently focuses on bug detection of equi-join queries. However, the idea of DSG and KQE can be extended to non-equal joins. The only challenge is how to generate and manage ground-truth results, whose sizes increase exponentially for non-equal joins. We leave it to our future work.

\section{Data-guided Schema and Query Generation (DSG)}

SQLancer~\cite{SQLancer} is a tool to automatically find logic bugs in the implementation of DBMS. It first creates a populated database with randomly generated tables. Afterwards, it randomly chooses SQL statements to create, modify, and delete data. Based on the randomly generated database, it adopts testing approaches such as Pivoted Query Synthesis (PQS)~\cite{rigger2020testing} to  detect logic bugs.
Note that SQLancer is designed for logic bug detection of single-table queries. We can extend it directly to multi-table queries by modifying its data and query generator, but many bugs are ignored in that case. For details, please refer to the experimental section.

\begin{figure*}[ht!]
    \centering
        \centering
        \includegraphics[scale=0.62]{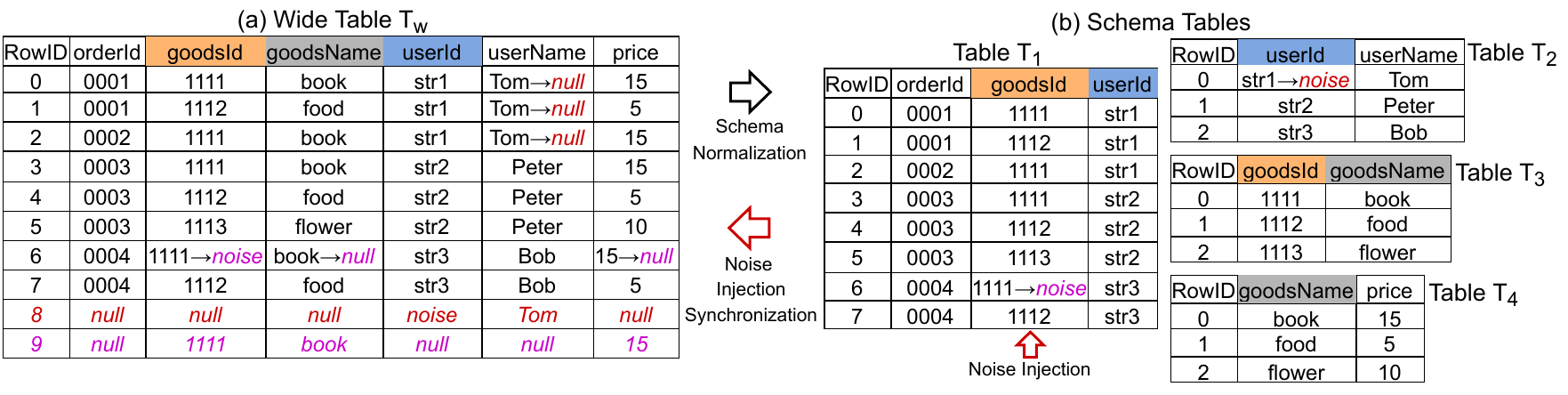}
         \vspace*{-0.2cm}
        \caption{Schema generation of the shopping order dataset. Data in black is the original dataset, and data in color is the noisy data which is injected in schema tables and then synchronized in the wide table.}
        \vspace{-0.2cm}
        \label{fig.schema}
\end{figure*}

To effectively support bug detection of multi-table join queries, we generate a test database as a wide table and leverage schema normalization to split the wide table into multiple tables with normal form. 
In addition, we propose an effective noise injection techniques to increase the probability of causing logic bugs with inconsistent database states. With the constructed database, join queries are generated via random walk on the schema graph and represented as abstract syntax trees. Finally,  we efficiently retrieve their ground-truth result sets for these generated queries, assisted by the proposed bitmap index.

\subsection{Schema Normalization}

We first generate a test database as a wide table $T_w$. $T_w$ can be constructed from either a real dataset (e.g., the KDD Cup dataset\footnote{https://archive.ics.uci.edu/ml/datasets/KDD+Cup+1998+Data} in the UCI machine learning repository is essentially a wide table) or a random database generator (e.g., TPC-H generator). For the latter case, we pick unbiased random samples from the fact table {\em lineitem}, and apply the primary-foreign key joins to merge it with the dimension tables to produce a wide table.

Given a wide table $T_w$, we apply schema normalization techniques to generate a testing database schema.
In the past literature, there have been numerous algorithms proposed for FD (functional dependency) discovery (such as TANE \cite{huhtala1999tane} and HYFD \cite{DBLP:conf/sigmod/PapenbrockN16}) and schema normalization~ \cite{DBLP:journals/tods/DiederichM88, DBLP:conf/edbt/PapenbrockN17}. The discovered FDs are used to transform the database schema into the 3NF. We directly use these data-driven schema normalization methods to generate our testing database schema. Note that we focus on 
FDs supported by the data,
not semantically correct ones, and hence, our schema generation is completely automatic. During the splitting process of $T_w$, 
we also maintain an explicit primary key $RowID$ for all generated tables in order to recover ground-truth results. Moreover, metadata about the implicit primary and foreign key relationships are also maintained.

\begin{example}
\vspace{-0.2cm}
We use an example of wide table in Figure \ref{fig.schema} to illustrate the idea. The FD discovery algorithm finds four valid FDs: \texttt{\{orderId, goodsId, userId  $\rightarrow$  goodsName, userName, price\}},
\texttt{\{goodsId  $\rightarrow$ goodsName,   price\}}, \texttt{\{goodsName  $\rightarrow$  price\}} and \texttt{\{userId $\rightarrow$ userName\}}.
The FDs are automatically selected for decomposing the wide table $T_w$, and the explicit $RowID$ columns are created.
The example table is decomposed into schema $S$ with four tables: $\{ T_1$(RowID, orderId, goodsId, userId),
$T_2$(RowID, userId, userName), $T_3$(RowID, goodsId, goodsName), $T_4$(RowID, goodsName, price)\}, where \{orderId, goodsId, userId\}, \{goodsId\}, \{goodsName\} and \{userId\} are implicit primary keys, and $\{T_1.userId \! \rightarrow \! T_2.userId\}$, $\{T_1.goodsId \! \rightarrow \! T_3.goodsId\}$ and $\{T_3.goodsName \! \rightarrow \! T_4.goodsName\}$ are implicit foreign key mappings.
\vspace{-0.2cm}
\end{example}

To facilitate join query synthesis and the ground-truth result generation, we also create a RowID mapping table $T_{RowIDMap}$, 
which defines a mapping relation $[RowID, T_i, row_j]$. Here, $row_j$ denotes the row id of table $T_i$. The mapping relation is the list of rows in the wide table $T_w$ which are split to create the $row_j$th row of table $T_i$. Based on the RowID map table, we build a join bitmap index which will be used to speed up the retrieval of ground-truth results. The bitmap index consists of $k$ bit arrays of size $n$, where $k$ and $n$ are the number of schema tables and data rows, respectively. Each row has been assigned a distinct $RowID$. For the bit array of value $T_j$, the $i$th bit is set to ``1'' if the $i$th record of table $T_w$ has produced some rows in Table $T_j$; otherwise, the $i$th bit is set to ``0''. If the table is too large and results in a sparse bitmap, we apply RLE (run-length encoding)-based technique, the WAH encoding \cite{DBLP:conf/ssdbm/WuOS02}, to compress consecutive sequences of ``0'' or ``1''.

\begin{figure}[h!]
  \centering
  \includegraphics[scale=0.6]{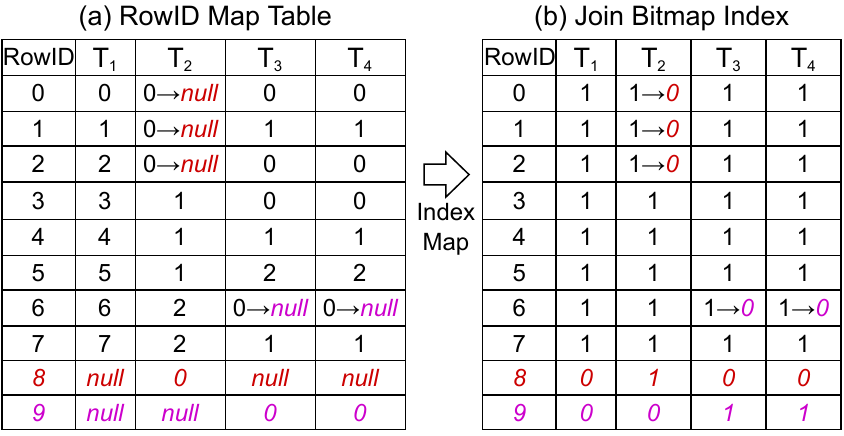}
  \caption{RowID map table $T_{RowIDMap}$ and the join bitmap index are built to retrieve the ground-truth of query joins. Data in color represents data updates after noise injection.}
  \vspace{-0.2cm}
  \label{fig.bitmap}
\end{figure}

\begin{example}
Figure \ref{fig.bitmap} illustrates examples of RowID map table and join bitmap index of the data in Figure \ref{fig.schema}.
The row with RowID $5$ of rowID map table records a split process, which splits the row with RowID $5$ in wide table $T_w$ to produce rows in table $T_1$, $T_2$, $T_3$ and $T_4$ with RowID $5$, $1$, $2$ and $2$ respectively.
The row with RowID $0$ of join bitmap index represents that tables ($T_1$, $T_3$, $T_4$) have rows split from the wide table with RowId $0$, while table $T_2$ has no corresponding row.
Data in color represents data updates after noise injection and will be discussed later.
\end{example}

\subsection{Noise Injection}
We first perform data cleaning task on our generated database by removing noisy data, which cause untraceable join results, e.g., null and boundary values on primary and foreign keys. 
Then, to facilitate the detection of logic bugs, we deliberately insert noises into the generated database $D_s$ to violate the FDs and primary-foreign key relationships. The injected noises produce traceable join results.
The idea of noise injection is to corrupt a small fraction ($\epsilon$) of the primary-foreign key relationship by replacing original values with (1) boundary values (e.g., for integer value and char(10) type, we replace the value with 65535 and `ZZZZZZZZZZ'), and (2) NULL values. For each primary and foreign key column, we randomly pick $\epsilon$ tuples to perform the value replacement. 
The produced noisy database $D_s'$ follows the same schema $S$ as that of $D_s$. 
When we inject noises, we guarantee that the values of injected noises are unique and do not violate the ground-truth results of normal data.

The introduction of noises violates the consistency between the generated tables and the original wide table $T_w$. Since our ground-truth result is recovered from the wide table, we need to update $T_w$ according to the injected noise so that the wide table and the noisy database become consistent.

Suppose a noise is introduced into column $col_k$ of the $row_j$th row in table $T_i$, we have two cases:

\textbf{Case 1:} 
If $col_k$ is the implicit primary key column, 
the affected rows in $T_w$ can be represented as 
$\bar{R} = RowMap(T_i, row_j)$, and 
the affected columns in $T_w$ can be represented as 
$\bar{C}=Fd(col_k)$, 
where $Fd(col_k)$ denotes the columns which are functional dependent on column $col_k$. Then,
cells in Table $T_w$ should be updated as follows:
\begin{equation}
\begin{aligned}
    insertion:&\
    T_w[N+1][col_k] = T_i[row_j][col_k],\\
    & T_w[N+1][c_k] =
    T_w[r][c_k] \mid r\in \bar{R}, \forall c_k\in \bar{C};\\
    update:\ & T_w[r_j][c_k] = NULL \mid \forall r_j\in \bar{R}, \forall c_k\in \bar{C}.
 \nonumber
\end{aligned}
\end{equation}
Here, $N$ denotes the number of tuples in $T_w$ and there involve two insertion operators and an update operator. The insertion operator creates a new tuple by copying the noisy data from $T_i$ and its function-determined values, leaving the remaining values as NULL. In the wide table, we can find multiple rows ($\forall r_j\in \bar{R}$) that can be functionally derived from the noisy row. We only need to copy one of them ($r\in \bar{R}$).
The update operator, on the other hand, modifies the rows of $T_w$ that relates to the $row_j$th row
of $T_i$ by tagging the corresponding column values as NULL, since the primary-foreign key joins are invalid.

\textbf{Case 2:} 
If the noise is created in the foreign key column of table $T_i$,
the tuples in Table $T_w$ should be updated as follows:
\begin{equation}
\begin{aligned}
   insertion: &  T_w[N+1][c_k]= T_w[r][c_k] \mid r\in \bar{R}, \forall c_k\in col_k\cup  \bar{C};\\
   update: & T_w[r_j][col_k] = T_i[row_j][col_k]\mid \forall r_j\in \bar{R},\\
    & T_w[r_j][c_k] = NULL \mid \forall r_j\in \bar{R}, \forall c_k\in\bar{C}.  \nonumber
\end{aligned}
\end{equation}
In case 2, we have an insertion and two update rules, respectively. This update process is explained in Example \ref{noise_example}.

\begin{figure}
  \centering
  \includegraphics[scale=0.42]{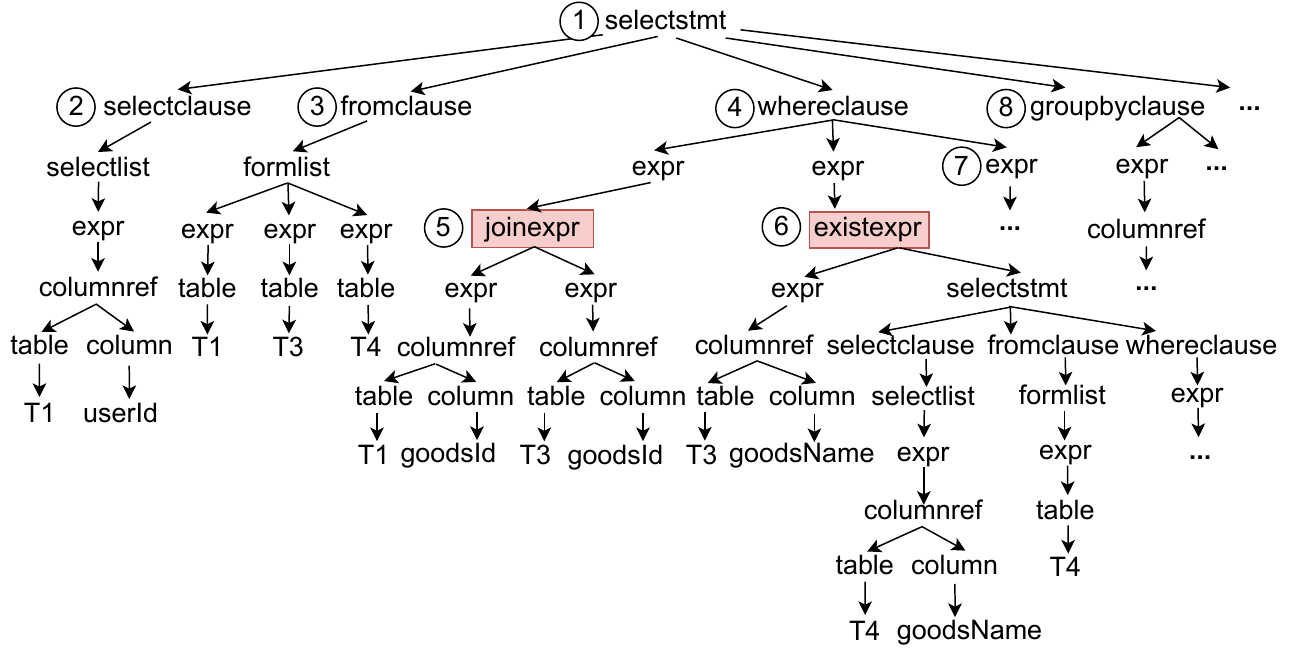}
  \vspace*{-0.5cm} 
  \caption{
  Example of join query generation. The join expressions are generated by random walk on the schema graph.}
  \label{fig.ast}
  \vspace*{-0.4cm} 
\end{figure}

After updating wide table $T_w$, we should adjust the RowID map table as well. We denote those columns in the RowID map table as $C_{dep}=\mathcal{T}(col_k\cup \bar{C})$, where $\mathcal{T}(\cdot)$ is the tables whose columns is a subset of the given columns.
\begin{equation}
\begin{aligned}
  insertion:\ &  T_{RowIDMap}[N+1][C_{dep}] = T_{RowIDMap}[r][C_{dep}] \mid r\in \bar{R};\\
  update:\ &  T_{RowIDMap}[r_j][C_{dep}] = NULL \mid \forall r_j\in \bar{R}. \nonumber
\end{aligned}
\end{equation}
The two rules are defined similarly to those for $T_w$ and we discard the details.

\begin{example}
\label{noise_example}
Figure \ref{fig.schema} illustrates the process of noisy injection. Initially, we inject noise into tables $T_1$ and $T_2$, which are highlighted with colored fonts.
To maintain a correct ground-truth result set, the corresponding rows of wide table $T_w$ also need to be synchronized according to the injected noise in Tables $T_1-T_4$. 

For table $T_2$, we inject a noisy value into the primary key $userId$ of the first tuple.
For $T_w$, we locate the rows corresponding to the noise via the RowID map table in Figure \ref{fig.bitmap}(a). The results are the rows in $T_w$ with RowID in $[0-2]$.
Using the insertion rule,
a new tuple (tuple 8) is created in $T_w$ to refer to the noisy tuple in $T_2$.
Similarly, based on the update rule, the tuples with RowID in $[0-2]$ need to change their {\em userName} to NULL values.

Similarly, table $T_1$ is polluted with a noisy value in its foreign key column $goodsId$.
We locate involved rows in wide table $T_w$ by mapping the noise through the RowID map table in Figure \ref{fig.bitmap}(a), which is the tuple in $T_w$ with RowID 6.
Because $goodsId$ is the primary key of $T_3$, when joining them together, the missing contents should be recovered.
Therefore, a new row with RowID 9 is created in $T_w$ to maintain contents of columns
$col_k\cup \bar{C}$.
On the other hand, the tuples with column $goodsId$ 
should be updated to noisy data, and the columns that can be functionally determined by $goodsId$ should be updated to NULL. 

Finally, the RowID map table should be updated by noises,
and the data in the join bitmap index which cannot be found by its RowID in the wide table should be set to 0.
\end{example}

\subsection{Join Query Generator}
Random walk-based workload generators have been adopted for
graph data exploration, such as RDF (resource description framework) search \cite{DBLP:conf/icde/BaganBCFLA17, DBLP:conf/semweb/AlucHOD14}.
We adopt a similar idea in our join query generation.
DSG generates abstract syntax trees (ASTs) up to a specified maximum depth to get the skeleton of syntax-correct SQL queries. 
The AST is a tree of objects that represents a {\em select} SQL statement. 
As Figure \ref{fig.ast} shows, the root of the tree is an object called selectstmt (node \textcircled{1}), which refers to a {\em select} SQL statement, and its child branches represent the {\em select}, {\em from} and {\em  where} clauses in the {\em select} statement.

To explore possible query space, we represent the generated database schema as a schema graph $G_s = {(V, E)}$ with nodes $v_i \in V$ and edges $(v_i, v_j) \in E$. $V$ can be further classified into two types of vertices, table vertex $V_t$ and column vertex $V_x$.
If  $v_i \in V_t$ and $v_j \in V_t$, the edge $(v_i, v_j)$ denotes that the corresponding two tables can be joined using primary-foreign key relationship.
If $v_i \in V_x$ and $v_j \in V_t$, the edge $(v_i, v_j)$ indicates that $v_i$ is a specific column of table $v_j$. Note that there will be only one edge per $v_i$ from $V_x$.

DSG adopts random walk on schema graph $G_s$ to select tables for queries, and uses those tables to generate join expressions.
This is a stochastic process, starting with a table vertex $v_i\in V_t$, with random variables $(W_{v_i}^{1}, W_{v_i}^{2},..., W_{v_i}^{k})$ such that $W_{v_i}^{k+1}$ is 
an edge chosen at random from the neighborhood of vertex $v_k$, where $v_k$ is the $k$th vertex reached by this stochastic process (i.e., the end point of edge $W_{v_i}^{k}$).
If the random walk process picks a (table-table) edge (say $(v_i, v_j)$ s.t. $v_i, v_j\in V_t$), a join relationship is obtained and we move to the new table vertex $v_j$. On the other hand, if the random walk process picks a (table-column) edge (say $(v_i, v_j)$ s.t. $v_i\in V_t$ and $v_j\in V_x$), random filters (i.e., a selection condition) are generated on the column $v_j$ and the random walk process continues from the preceding table vertex $v_i$ (but now excluding $v_j$).

After the join expressions are generated by random walk on schema graph $G_s$, DSG randomly generates other expressions based on the {\em join} clauses.
Generating these expressions is implemented similarly to RAGS~\cite{DBLP:conf/vldb/Slutz98} and SQLSmith~\cite{SQLSmith}.
Note that we support sub-query inside the IN/Exist expressions of the {\em where} clause.

\begin{example}
Figure \ref{fig.ast} depicts a running example of join query generation. 
The random walk process selects three tables ($T_1$, $T_3$, $T_4$) to join using the join conditions involving columns $goodsId$ and $goodsName$
(e.g., the nodes \textcircled{5} and \textcircled{6} in Figure \ref{fig.ast}).
The join type of node \textcircled{5} can be inner/outer/cross join, node \textcircled{6} can be semi/anti-join.
As Figure \ref{fig.ast} shows, the table expressions of the {\em from} clause (node \textcircled{3}) are set to the tables 
which are involved in the join processing.
Then, {\em select} clause (node \textcircled{2}) and {\em where} clause (node \textcircled{7}) are randomly constructed for tables of {\em from} clause and the random walk results (i.e., the columns and their types).
And the aggregation operators are also supported (node \textcircled{8}).
Finally, we transform the AST back to a SQL statement.
\end{example}

\subsection{Ground-truth Result Generation}

Given a join query generated by random walk on the schema graph, DSG can efficiently retrieve the ground-truth result from the wide table $T_w$. In the following, we first propose the strategies to support different join operators. Then, we present the procedure for ground-truth result generation.

DSG supports seven types of join operators, including inner join, left/right/full outer join, cross join and semi/anti-join. The ground-truth for the join bitmap index of the join results (which is a table), for each join type, is summarized in Table~\ref{tab.groundtruth}. For an inner join, the ground-truth of the RowID join bitmap index is $Bit({T_i}) \wedge Bit({T_{i+1}})$, where $Bit({T_i})$ is the RowID join bitmap index of Table $T_i$ (i.e., the column for $T_i$ in Figure \ref{fig.bitmap}(b)) and $\wedge$ is the bitwise logical AND operator.
For a left outer join, the ground-truth bitmap is  $Bit({T_i})$.
The same rule is also applied to the right outer join.
For a full outer join, the ground-truth bitmap is $Bit({T_i}) \vee Bit({T_{i+1}})$
and $\vee$ is the bitwise logical OR operator.
For the cross join, the full result set cannot be recovered via $T_w$, and hence, we resort to verifying a subset of the result set. In other words, the database system must return a result set containing all the ground-truth answers from $T_w$.
For a semi-join, since we stick to the primary-foreign key joins, the ground-truth bitmap is $Bit({T_i}) \wedge Bit({T_{i+1}})$. 
Finally, DSG also supports the anti-join, whose ground-truth bitmap is denoted as $Bit({T_i}) \wedge\! \sim\! Bit({T_{i+1}})$.

Based on the rules in Table \ref{tab.groundtruth}, we can generate a complete join bitmap index for queries involving multiple joins via bitwise AND of the corresponding RowID join bitmap index.
For the example query in Figure \ref{fig.ast}, 
if the join type of node \textcircled{5} is inner join and node \textcircled{6} is anti-join,
the join bitmap of the query is 
\begin{equation}
    Bit(T_1\! \Join_1 T_3\! \Join_2\! T_4) = Bit({T_1}) \wedge Bit({T_3}) \wedge \sim Bit({T_4}),
    \label{eq.joinbit}
\end{equation}
where $\Join_1$ denotes the inner join, and $\Join_2$ is the anti-join.
To reduce the overhead of bitmap calculation, the jump intersection algorithm is adopted
to avoid unnecessary intersections of the RowID list.
We rank the bitmaps based on their sparsity and conduct the intersection, starting from the most sparse bitmap.

After obtaining the ground-truth bitmap, we apply it to the wide table $T_w$ to retrieve the tuples participating in the join query. 
Duplicated tuples are removed based on the new primary key of the tuples.
Then, DSG also executes the generated filters and projections defined in the AST. In this way, we can obtain the ground-truth result for a specific random join query on our generated schema graph.

\begin{example}
Now consider the query of "SELECT price FROM T3 INNER JOIN T4 WHERE T3.goodsName = T4.goodsName AND T3.goodsName = `flower'".
Firstly, using the rules in Table \ref{tab.groundtruth}, we get the join bitmap of the query: $Bit(T3)\wedge Bit(T4)$, and retrieve redundant data with RowID $\{0-5,7,9\}$ in wide table $T_w$ by using join bitmap index in Figure \ref{fig.bitmap}(b).
Then, the duplicates are removed based on the primary key $goodsId$. Hence, the remaining results of the query include tuples with RowID $\{0, 1, 5\}$ in the wide table $T_w$.
Finally, filters and projections are also applied, and the ground truth of the query is $``10"$.
\end{example}

    \begin{table}
  \centering
    \caption{The ground-truth bitmap for each join type.}
    \label{tab.groundtruth}
    \begin{tabular}{|c|c|c|}
    \hline
     Type of Join ($T_i \Join_i  T_{i+1}$)       & Verification & $Bit({T_i \Join_i  T_{i+1}})$ \\ \hline
     Inner Join    & FullSet & $Bit({T_i}) \wedge  Bit({T_{i+1}})$ \\\hline
     Left Outer Join  & FullSet & $Bit({T_i})$ \\\hline
     Right Outer Join & FullSet & $Bit({T_{i+1}})$ \\\hline
     Full Outer Join & FullSet & $Bit({T_i} \vee Bit({T_{i+1}})$ \\ \hline
     Cross Join       & SubSet  & $Bit({T_i}) \wedge  Bit({T_{i+1}})$ \\\hline
     Semi Join        & FullSet & $Bit({T_i}) \wedge  Bit({T_{i+1}})$\\\hline
     Anti Join        & FullSet & $Bit({T_i}) \wedge  \sim Bit({T_{i+1}})$\\\hline 
    \end{tabular}
    \end{table}

\section{Knowledge-guided Query Space Exploration (KQE)}\label{sec:kqe}

\begin{figure}[h!]
  \centering
 \includegraphics[scale=0.75]{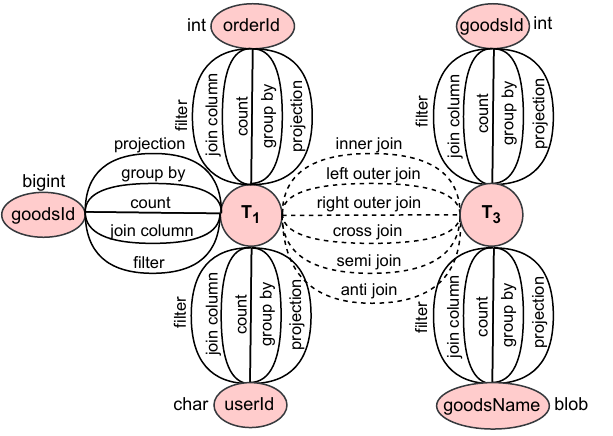}
   \caption{Plan-iterative graph. Queries can be mapped to sub-graphs in the plan-iterative graph.}
  \label{fig.schemagraph}
\end{figure}

Given a schema, the search space of valid join queries is huge and it is infeasible and ineffective to enumerate all possibilities. 
To represent the search space, we first extend the schema graph $G_s$ to a plan-iterative graph
$G=(V, E')$. Vertices in $G$ are classified into two types, table vertex $v_t$ with label ``table'' and column vertex $v_c$ with label $type$, indicating data type of the column. 
Based on the vertex type, edges are also split into two categories: edges between two table vertexes $(v_t, v_t')$ and edges between table vertexes and column vertexes $(v_t, v_c)$. The labels of $(v_t, v_t’)$ and $(v_t, v_c)$ denote the join type and the relational operator applied to the column, respectively. 
An example of plan-iterative graph is shown in Figure \ref{fig.schemagraph}. 
If $m$ join operators are supported, we will create $m$ edges between two tables with primary-foreign key relationship. 
If we pick ``filter'' on the edge between $T_1$ and {\em userId}, we will generate a predicate on {\em userId} in the generated query. 
To simplify the representation,
we define a label function $f_l$ which maps a vertex $v \in V$ or an edge $e \in E$ to its label.

With the plan-iterative graph $G_s$, each generated join query can be mapped to a sub-graph in $G_s$.
If the corresponding sub-graphs of two generated queries are isomorphic, this implies that they share the same query structure and incur redundant examination overhead in the query space. In other words, using either one of the isomorphic queries is sufficient for logic bug detection. 
Therefore, we enforce the query generation algorithm with a constraint that a newly generated query graph $G_q$ should not be isomorphic with some existing query graphs $G_d$.
In the following, we formally define sub-graph isomorphism and isomorphic set.
\begin{definition}
Given $G_q=(V,E)$ and $G_d=(V',E')$, a sub-graph isomorphism is an injection function $f_{iso}$ from $V$ to $V'$ such that $\forall v \in V$, $f_l(v)=f_l(f_{iso}(v))$;  $\forall e(v, v') \in E$, $e(f_{iso}(v),f_{iso}(v'))\in E'$ and $f_l(e(v,v'))=f_l(e(f_{iso}(v),f_{iso}(v')))$.
\end{definition}
\begin{definition}
Given a plan-iterative graph $G$, an isomorphic set $S$ contains a set of sub-graphs of $G$ and has the following properties: 1) $\forall G_i, G_j\in S$, $G_i$ and $G_j$ are isomorphic; 2) we cannot find another sub-graph $G_x$ of $G$ having $G_x\notin S$ and $\exists G_i\in S$, $G_x$ and $G_i$ are isomorphic.
\end{definition}

Unfortunately, we cannot directly evaluate the exact graph isomorphism for every generated query because it has been proven that determining the sub-graph existence in a graph (i.e., sub-graph isomorphism matching) is NP-complete \cite{bodirsky2015graph, DBLP:books/fm/GareyJ79}.
Recently, learning-based models have been proposed for sub-graph analysis tasks. For examples, NeuroMatch \cite{DBLP:journals/corr/abs-2007-03092} and LMKG \cite{DBLP:conf/edbt/DavitkovaG022} are developed for the sub-graph isomorphism problem. These algorithms perceive the presence or absence of a sub-graph problem as a binary classification problem.
To support more general sub-graph isomorphism search, GNNs (Graph Neural Networks) have been adopted by  \cite{DBLP:journals/pvldb/DuongHYWNA21} and \cite{ DBLP:journals/pvldb/YangFO0021}.

\begin{figure}[b!]
  \centering
\vspace{-0.2cm} 
 \includegraphics[scale=0.85]{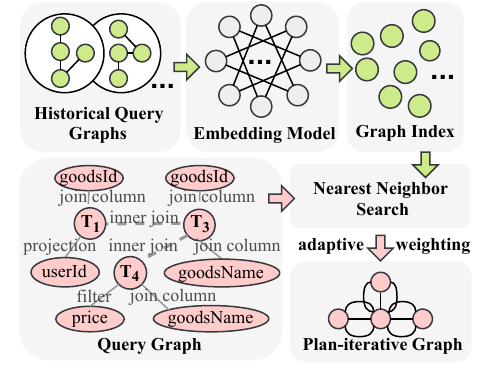}
   \vspace{-0.2cm} 
   \caption{Running example of knowledge-guided query space exploration.}
  \label{fig.kqe}
   \vspace{-0.2cm} 
\end{figure}

Motivated by learning-based approaches for sub-graph isomorphism search, we extend our random walk scheme in Section 3.3 with an adaptive weighting strategy. In order to avoid repeatedly exploring similar graph structures, we adjust the probabilities of random walks based on the exploration history. Figure \ref{fig.kqe} illustrates our idea.
To support the approximate evaluation of subgraph isomorphism, KQE builds a graph index {\em GI}.
Given a query $q$ and its corresponding sub-graph $G_q$ from the plan-iterative graph,
{\em GI} first applies the similarity-oriented graph
embedding approach~\cite{DBLP:journals/pvldb/DuongHYWNA21} to generate a unique high dimensional embedding $E(G_q)$ for $q$. If two sub-graphs are isomorphic or structurally similar, the cosine distance of their embeddings is expected to be below a threshold.
Then, {\em GI} applies the HD-Index \cite{DBLP:journals/pvldb/AroraSK018} to support approximate KNN (K-Nearest Neighbor) search in the high-dimensional space.

We define the coverage score of a generated query graph $G_q$ wrt historical sub-graphs (i.e., query graphs of queries that had already been explored): 
\begin{equation}
    coverage(G_q)=\frac{1}{k}\sum_{i=1}^{k}cosine\_similarity(E(G_q), E(G_i)),
\end{equation}
where $G_i$ is the $i$-th nearest neighbor of the query graph $G_q$ returned by {\em GI}. A higher coverage value indicates that similar graph structures have been explored and hence, this indicator allows us to avoid repeatedly generating similar queries. 

\begin{algorithm}[t]
\caption{AdaptiveRandomWalk ($G, v, l, GI$)}  
\label{alg.randomwark}
\begin{algorithmic}[1]
\Require plan-iterative graph $G$, start node $v$, 
maximum walk length $l$, graph index {\em GI}.
\Ensure walk $W_{v}$.
\State Initialize walk $W_{v}$ to $[v]$, $\pi_{e_x}=1$
\For{$k=1$ to $randomInt(l)$}
    \State{$v_k = W_{v}$.last\_vertex}
    \State{$G_q = getGraph(W_{v})$}
    \State{$E_{k}=getNextEdges(v_k, G)$}
    \For{$e_x$ in $E_k$}
      \State{$G_q' = G_q$ extend by $e_x, v_x$ (i.e., $e_x=(v_k, v_x)$)}
      \State{obtain $\pi_{e_x}$ by GI}
    \EndFor
    \If{$\pi_{e_x} < \pi_{end}$}
       \Break
    \EndIf
    \State{$e_{k+1}=AliasSample(E_{k}, \pi_{e_x})$}
    \State{$v_{k+1}=getNode(v_k, e_{k+1})$}
    \State{append $[e_{k+1}, v_{k+1}]$ to $W_{v}$}
    \State{$\pi_{end} = \pi_{e_{k+1}}$}
\EndFor
\State \Return $\ W_{v}$
\end{algorithmic}  
\end{algorithm} 

To guide the query generator to explore more diversified search spaces, KQE performs the random walk according to the adaptive weights of the next possible edges.
Consider a random walk that follows a path
${P_q=(v_0, e_1, v_1, ...,e_k, v_k)}$ 
to generate a sub-graph $G_q$, with $v_k$ being the current vertex that is accessed, KQE decides its next step by evaluating the transition probability $\pi_{e_x}$ for each edge $e_x$ connecting with vertex $v_k$. A possible next query graph $G_q'$ can be created by adding edge $e_x$ and the corresponding vertex $v_x$ (i.e., $e_x=(v_k, v_x)$) to $G_q$ (i.e., ${P_{q'}=(v_0, e_1, v_1, ...,e_k, v_k, e_x, v_x)}$).
The transition probability on edge $e_x$ is set to
\begin{equation}
    \pi_{e_x}=\frac{1}{coverage(G_q')+1}.
    \label{formula:probability}
\end{equation}
We also establish a termination mechanism with a probability $\pi_{end}$, when the scores of all next possible graphs are less than the current query graph, the graph expansion is stopped. 
The pseudocode for the adaptive random walk on plan-iterative graph $G$ is given in Algorithm \ref{alg.randomwark}. 
At every step of the walk, alias sampling is conducted based on the transition probability $\pi_{e_x}$. 
Sampling of nodes while simulating the random walk can be done efficiently in $O(1)$ time complexity using alias sampling.

The existence of KQE allows us to perform parallel query space exploration, where a central server hosts the graph index and applies the adaptive random walk approach. When a new query is generated, the server disseminates it to a random client, which maintains a replica of the database and hosts an individual DSG process. This strategy effectively improves the efficiency of database debugging, and the only bottleneck is the synchronization cost of the KQE on the server side.

\section{Experiment}
We evaluate the TQS on multiple DBMSs and 
specifically, our goal is to answer the following questions:
\begin{itemize}
\item Can the TQS detect logic errors of implementations of multi-table queries from real-world production-level DBMSs? (section \ref{sec.bugs})
\item Can the TQS outperform state-of-the-art testing tools? (section \ref{sec.comparison})
\item What are the main roles of two core modules (namely DSG and KQE) of TQS?
(section \ref{sec.ablation})
\end{itemize}

\textbf{Tested DBMSs.}
In our evaluation, we consider three open-source DBMSs designed for different purposes (see Table \ref{tab.dbms})
to demonstrate the generality of TQS.  
According to the DB-Engine's Ranking \cite{dbengines}, the Stack Overflow's annual Developer Survey \cite{stackover}, and GitHub, these DBMSs are among the most popular and widely-used DBMSs.
We also run TQS against our own cloud-native database system, which is anonymized as X-DB.
X-DB is designed to run on elastic computation and storage resources with high scalability and concurrency support.
All selected DBMSs support hints for their optimizers to intentionally change the physical plans \cite{MySQLhints, mariadbhints, TiDBhints}.

We have evaluated TQS with both randomly generated TPC-H data and the dataset from UCI machine learning repository\footnote{https://archive.ics.uci.edu/ml/datasets/KDD+Cup+1998+Data}. 
However, since TPC-H dataset follows uniform distribution and has a simple schema, all reported bugs on TPC-H dataset have been covered by the UCI dataset. Therefore, we only show the results on the UCI data.
As TQS runs, it continuously reports bugs and to be fair for all DBMSs, we only report the results for the first 24 hours.
All DBMSs are run with default configurations and compilation options.

\textbf{Baseline approaches.}
We compare TQS with the 
SQLancer\footnote{https://github.com/sqlancer/sqlancer} which is the start-of-the-art approach to detecting logic bugs in databases.
SQLancer is not designed to test multi-table queries. However, it can be tailored for
multi-table queries by artificially generating queries and tuples across more than one tables. Similar to the single table case, all queries and tuples are randomly generated.
We use three methods in SQLancer as our baselines.
The first one is PQS \cite{rigger2020testing}, which constructs queries to fetch a randomly selected tuple from a table.
The tested DBMS may contain a bug if it fails to fetch that tuple.
The second one is TLP \cite{rigger2020testing}, which decomposes a query into three partitioning queries, each of which computes its result on that tuple.
The third one is NoRec \cite{rigger2020detecting}, which targets at logic bugs generated by the optimization process in DBMS. 
It performs a comparison between the results of randomly-generated queries and their rewritten ones.

\textbf{DBMS versions}. 
Note that DBMS bugs will be fixed once reported to the community. In our experiments,
we report the results of the latest released versions during our testing, namely, MySQL 8.0.28, MariaDB 10.8.2, TiDB 5.4.0 and X-DB beta 8.0.18 respectively.

\begin{table}[t]
  \centering
    \caption{We tested a diverse set of popular and emerging DBMS. All numbers are the lasted as of July 2022.}
    \vspace{-0.2cm}
    \label{tab.dbms}
    \begin{tabular}{|c|c|c|c|c|c|}
    \hline
     \multirow{3}{*}{DBMS}  & \multicolumn{3}{c|}{Popularity Rank} & \multirow{3}{*}{LOC}  & \multirow{3}{*}{\makecell[c]{First \\Release}}  \\ \cline{2-4}
          & \makecell[c]{DB-\\Engines} & \makecell[c]{Stack\\ Overflow} & \makecell[c]{Github\\ Stars} & & \\ \hline
        MySQL   & 2 & 1 & 8.0k & 3.8M & 1995 \\ \hline
        MariaDB  & 12 & 7 & 4.3k & 3.6M & 2009  \\ \hline
        TiDB    & 96 & -- & 31.8k & 0.8M & 2017 \\ \hline
    \end{tabular}
     \vspace{-0.2cm}
\end{table}

\textbf{Setup.} All experiments are conducted on our in-house server equipped with Intel Xeon CPU E5-2682 (2.50GHz) 16 cores and 128GB memory. For our cloud-native database X-DB, we run an instance on our public cloud with similar computation capability to our in-house server.

\begin{table*}
  \centering
    \caption{Detected bugs. TQS found 
    20 bug types: 
    7 from MySQL, 
    5 from MariaDB, 5 from TiDB and 3 from X-DB.}
     \vspace{-0.2cm} 
    \label{tab.totalbugs}
    \begin{tabular}{|c|c|c|c|l|}
    \hline
     Database & ID & Status & Severity & Description    \\
    \hline
    \multirow{7}{*}{\makecell{MySQL\\ 8.0.28}} 
             & 1  & Fixed & S1 (Critical) & Semi-join gives wrong results.\\ \cline{2-5}
              & 2  & Fixed & S2 (Serious) & Incorrect inner hash join when using materialization strategy.\\ \cline{2-5}
              & 3  & Verified & S2 (Serious) & Incorrect semi-join execution results in unknown data.\\ \cline{2-5}
              & 4  & Verified & S2 (Serious) & Incorrect left hash join with subquery in condition.\\ \cline{2-5}
              & 5  & Verified & S2 (Serious) & Incorrect nested loop antijoin when using materialization strategy.\\ \cline{2-5}
              & 6  & Fixed & S2 (Serious) & Bad caching of converted constants in NULL-safe comparison.\\ \cline{2-5}
              & 7  & Verified & S2 (Serious) & Incorrect hash join with materialized subquery.\\
            \hline
    \multirow{5}{*}{\makecell{MariaDB\\10.8.2}} 
              & 8 & Verified & Major & Incorrect join execution by not allowing BKA and BKAH join algorithms. \\ \cline{2-5}
              & 9 & Verified & Major & Incorrect join execution by not allowing BNLH and BKAH join algorithms. \\ \cline{2-5}
              & 10 & Verified & Major & Incorrect join execution when controlling outer join operations. \\ \cline{2-5}
              & 11 & Verified & Major & Incorrect join execution by limiting the usage of the join buffers. \\ \cline{2-5}
              & 12 & Verified & Major & Incorrect join execution when controlling join cache.  \\ \hline
    \multirow{5}{*}{\makecell{TiDB\\ 5.4.0}} 
              & 13  & Fixed & Critical & Incorrect Merge Join Execution when transforming hash join to merge join.\\ \cline{2-5}
              & 14  & Fixed & Critical  & Merge Join executed incorrect resultset which missed -0.\\ \cline{2-5}
              & 15  & Fixed & Critical & Merge Join executed an incorrect resultset which returned an empty resultset.\\ \cline{2-5}
              & 16  & Fixed & Critical & Merge Join executed an incorrect resultset which returned NULL.\\ \cline{2-5}
              & 17  & Fixed & Critical & Merge Join executed an incorrect resultset which missed rows.\\ 
            \hline
    \multirow{3}{*}{\makecell{X-DB\\ 8.0.18}} 
              & 18  & Fixed & 
              2 (High)
              & Left join convert to inner join returns wrong result sets. \\ \cline{2-5}
              & 19  & Fixed & 
              2 (High)
              & Hash join returns wrong result sets.
              \\ \cline{2-5} 
              & 20  & Verified & 
              2 (High)
              & Incorrect semi-join with materialize execution. \\ 
            \hline
    \end{tabular}
    \vspace{-0.2cm} 
    \end{table*}

\subsection{Overview and Showcase of Bug Reports } \label{sec.bugs}

TQS has successfully detected 
115
bugs from tested DBMSs within 24 hours,
including 
31
bugs in MySQL, 
30
bugs in MariaDB, 31 bugs in TiDB, and 23 bugs in X-DB respectively.  However, some bugs may be caused by the same relational algebras. Hence,
once finding a possible bug, it is essential to produce a minimal test case with C-Reduce\footnote{http://embed.cs.utah.edu/creduce/} before reporting the bug, to save the DBMS developers' time and effort. 
Then, we submit all bugs and the corresponding test cases to developer communities and received their positive feedbacks.
Through such root cause analysis, there are totally 
7,5,
5 and 3 types of bugs in MySQL, MariaDB, TiDB and X-DB respectively.
Table \ref{tab.totalbugs} shows the details of the identified bug types by DBMS developers.

Among all 20
types of bugs, some result in code fixes (8 reports), documentation fixes (2 reports), or are confirmed by the developers (10
reports). In other words,
each bug is previously unknown and has a unique fix associated with it, or is confirmed by the developers to be a unique bug.

\textbf{Severity levels.} 
For the tested DBMSs, bugs are assigned a severity level by the DBMS developers. 
6 bugs were classified as Critical, 
6 bugs as Serious, 
5 bugs as Major and 3 bugs as High.
In what follows, we briefly explain those bugs.

\subsubsection{Bugs in MySQL}
A large portion of bugs in MySQL involves semi-join and sub-query execution.
Listing \ref{listing:mysql_semijoin} shows an example query resulting in the incorrect semi-join execution. 

\begin{lstlisting}[language=SQL,
        caption={MySQL's incorrect semi-join execution.},
        label=listing:mysql_semijoin]
CREATE TABLE t0(
  c0 DECIMAL ZEROFILL COLUMN_FORMAT DEFAULT);

INSERT HIGH_PRIORITY INTO t0(c0) VALUES(NULL), (2000-09-06), (NULL);
INSERT INTO t0(c0) VALUES(NULL);
INSERT DELAYED INTO t0(c0) VALUES(2016-02-18);

query: 
SELECT t0.c0 FROM t0 WHERE t0.c0 IN (SELECT t0.c0 FROM t0 WHERE (t0.c0 NOT IN (SELECT t0.c0 FROM t0 WHERE t0.c0)) = (t0.c0));

ground truth: 
Empty set

transformed query: 
SELECT t0.c0 FROM t0 WHERE t0.c0 IN (SELECT /*+ semijoin()*/ t0.c0 FROM t0 WHERE (t0.c0 NOT IN (SELECT t0.c0 FROM t0 WHERE t0.c0))=t0.c0);

result:
+------------+
| c0         |
+------------+
| 0000001985 |
| 0000001996 |
+------------+

transformed query: 
SELECT t0.c0 FROM t0 WHERE t0.c0 IN (SELECT/*+ no_semijoin()*/ t0.c0 FROM t0 WHERE (t0.c0 NOT IN (SELECT t0.c0 FROM t0 WHERE t0.c0))=t0.c0);

result:
Empty set
\end{lstlisting}

By analyzing query plans of the above query, we find that hash join with semi-join produces incorrect results when using materialization technique for optimization.
MySQL community responds to us that there is a documented fix in the changelog of MySQL 8.0.30. It explains that incorrect results may be generated from execution of a semi-join with materialization, when the WHERE clause has an equal condition. In some cases, such as when the equal condition is denoted as an IN or NOT IN expression, the equality is neither pushed down for materialization, nor evaluated as part of the semi-join. This could also cause issues with inner hash joins.

\begin{lstlisting}[language=SQL,
        caption={
        MySQL's incorrect left hash join execution.},
        label=listing:mysql_hashjoin]
CREATE TABLE t0 (
  c0 text NOT NULL,
  primary key (c0));

CREATE TABLE t1 (
  c0 tinyint(3) unsigned zerofill,
  c1 varchar(15) NOT NULL,
  primary key (c0),
  key t1_fk1 (c1),
  constraint t1_ibfk_1 foreign key (c1) 
  references t0(c0));

query:
SELECT t1.c0 FROM t1 LEFT OUTER JOIN t0 ON t1.c1 = t0.c0 WHERE (t1.c1 IN (SELECT t0.c0 FROM t0 )) OR (t0.c0);

ground truth:
+------+
|  c0  |
+------+
| NULL |
| NULL |
+------+

\end{lstlisting}
\lstset{style=mystyle}

We also test with foreign key constraints.
Listing \ref{listing:mysql_hashjoin} illustrates a bug when trying to optimize a left hash join using subquery\_to\_de-rived condition. MySQL retrieves an additional ``NULL'' value.

\subsubsection{Bugs in MariaDB}
Different from MySQL, many bugs were reported on the nested loops join and hash join in MariaDB. 
Listing \ref{listing:mariadb_bnl} shows an example bug, which is produced by transforming block nested loop hash join to block nested loop join in MariaDB. Incorrect result set is obtained, as 
when MariaDB executes the join, data are mistakenly changed to empty.

\begin{lstlisting}[language=SQL,
        caption={MariaDB's incorrect loop join execution.},
        label=listing:mariadb_bnl]
CREATE TABLE t1 (
  c0 varchar(100) NOT NULL, 
  KEY ic1 (c0));

CREATE TABLE t2 (
  c0 varchar(100) NOT NULL);
  
query: 
SELECT t2.c0 FROM t2 RIGHT OUTER JOIN t1 ON t1.c0=t2.c0;

ground truth:
+------+
| c0   |
+------+
| NULL |
| NULL |
+------+

transformed query: 
SET optimizer_switch='join_cache_hashed=off';
SELECT t2.c0 FROM t2 RIGHT OUTER JOIN t1 ON t1.c0=t2.c0;

result:
+------+
|  c0  |
+------+
|      |
| NULL |
+------+
\end{lstlisting}
\vspace{-0.1cm}

Listing \ref{listing:mariadb_cross} shows another bug when transforming 
batch key access join to block nested loop join. 
When MariaDB executes the join, the optimizer mistakenly changes the null value to empty.

\begin{lstlisting}[language=SQL,
        caption={MariaDB's incorrect index join execution.},
        label=listing:mariadb_cross]
CREATE TABLE t1 (
  c0 bigint(20) DEFAULT NULL);

CREATE TABLE t2 (
  c0 double NOT NULL, 
  c1 varchar(100) NOT NULL, 
  PRIMARY KEY (c1));

CREATE TABLE t3 (
  c0 mediumint(9) NOT NULL, 
  c1 tinyint(1) NOT NULL);

CREATE TABLE t4 ( 
  c0 double NOT NULL, 
  c1 varchar(100) NOT NULL, 
  PRIMARY KEY (c1));

query:
SELECT t3.c0 FROM t3 RIGHT OUTER JOIN t4 ON t3.c1 = t4.c1 JOIN t2 ON t2.c1 = t4.c1 CROSS JOIN t1;

ground truth:
+------+
|  c0  |
+------+
| NULL |
| NULL |
+------+

transformed query: 
SET optimizer_switch='join_cache_bka=off';
SELECT t3.c0 FROM t3 RIGHT OUTER JOIN t4 ON t3.c1 = t4.c1 JOIN t2 ON t2.c1 = t4.c1 CROSS JOIN t1;

result:
Empty set     
\end{lstlisting}

\begin{figure*}
    \centering
    \vspace{-0.2cm}
    \subcaptionbox{MySQL diversity}{
      \begin{minipage}[c]{.23\linewidth}
        \centering
        \includegraphics[scale=0.35]{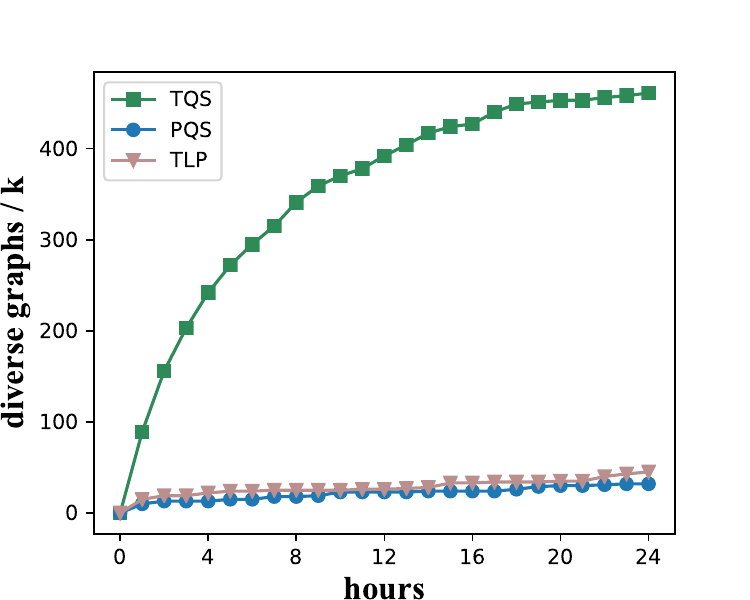}
      \end{minipage}
      \vspace{-0.1cm}
    }
    \subcaptionbox{MariaDB diversity}{
      \begin{minipage}[c]{.23\linewidth}
        \centering
        \includegraphics[scale=0.35]{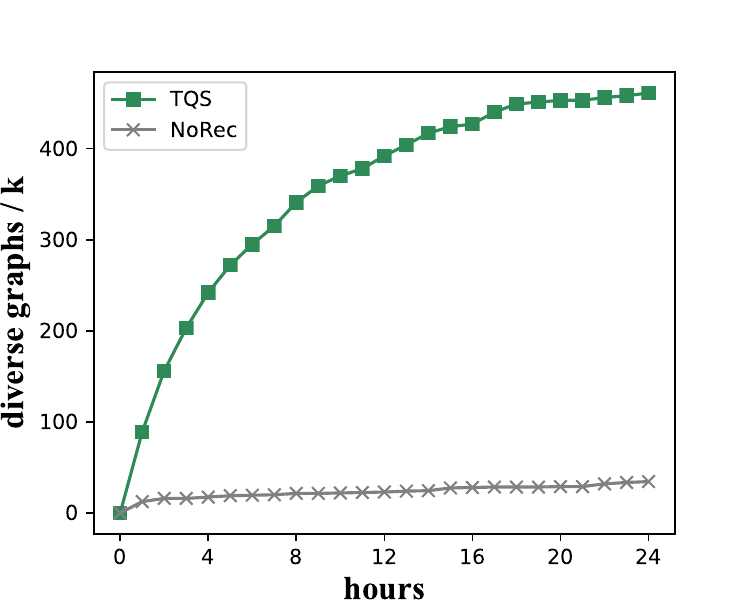}
      \end{minipage}
      \vspace{-0.1cm}
    }
    \subcaptionbox{TiDB diversity}{
      \begin{minipage}[c]{.23\linewidth}
        \centering
        \includegraphics[scale=0.35]{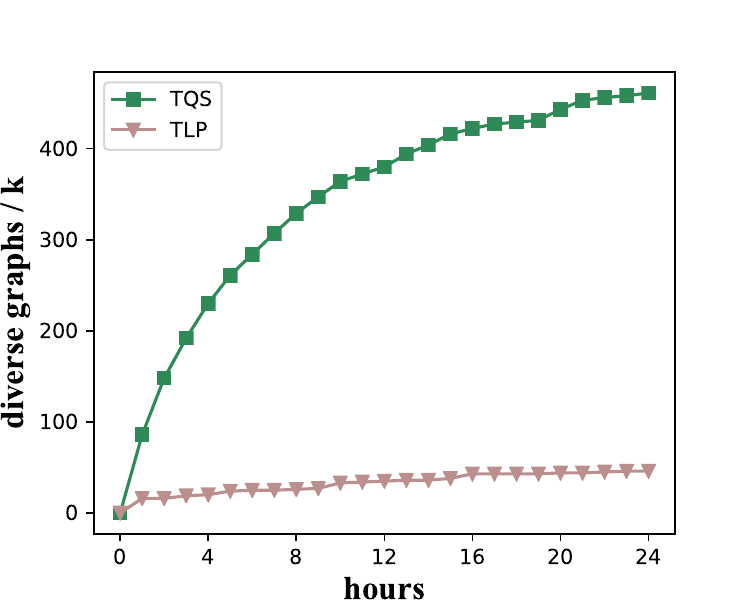}
      \end{minipage}
      \vspace{-0.1cm}
    }
    \subcaptionbox{X-DB diversity}{
      \begin{minipage}[c]{.23\linewidth}
        \centering
        \includegraphics[scale=0.35]{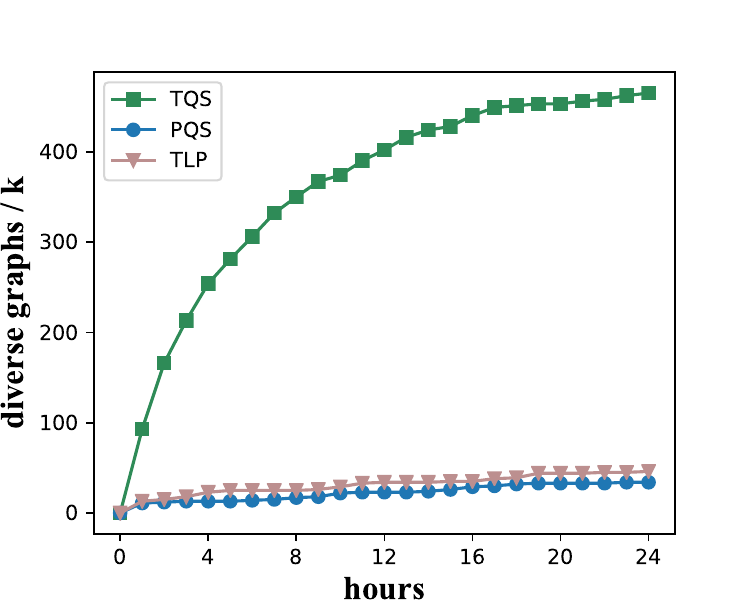}
      \end{minipage}
      \vspace{-0.1cm}
    }
    \subcaptionbox{MySQL efficiency}{
      \begin{minipage}[c]{.23\linewidth}
        \centering
        \includegraphics[scale=0.35]{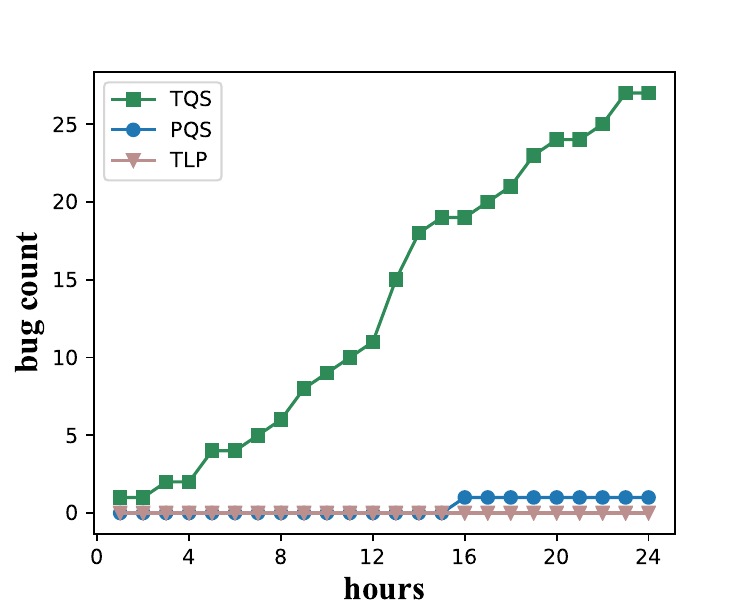}
      \end{minipage}
      \vspace{-0.1cm}
    }
    \subcaptionbox{MariaDB efficiency}{
      \begin{minipage}[c]{.23\linewidth}
        \centering
        \includegraphics[scale=0.35]{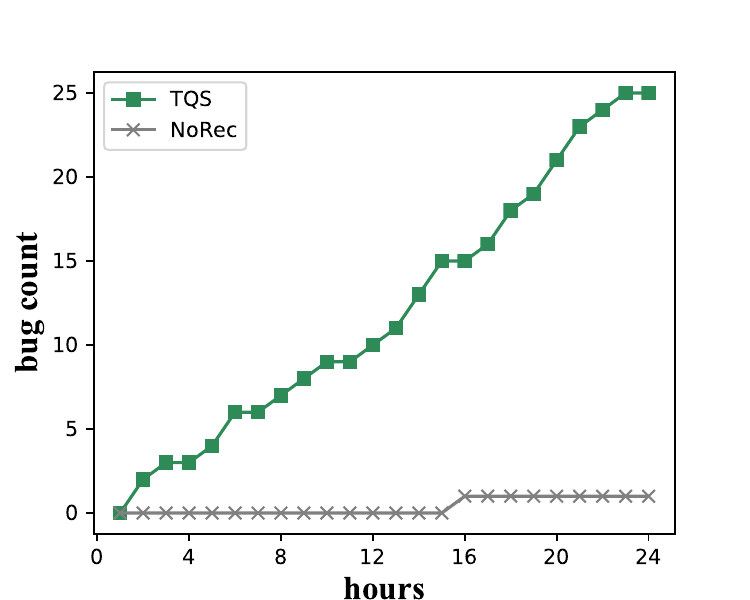}
      \end{minipage}
      \vspace{-0.1cm}
    }
    \subcaptionbox{TiDB efficiency}{
      \begin{minipage}[c]{.23\linewidth}
        \centering
        \includegraphics[scale=0.35]{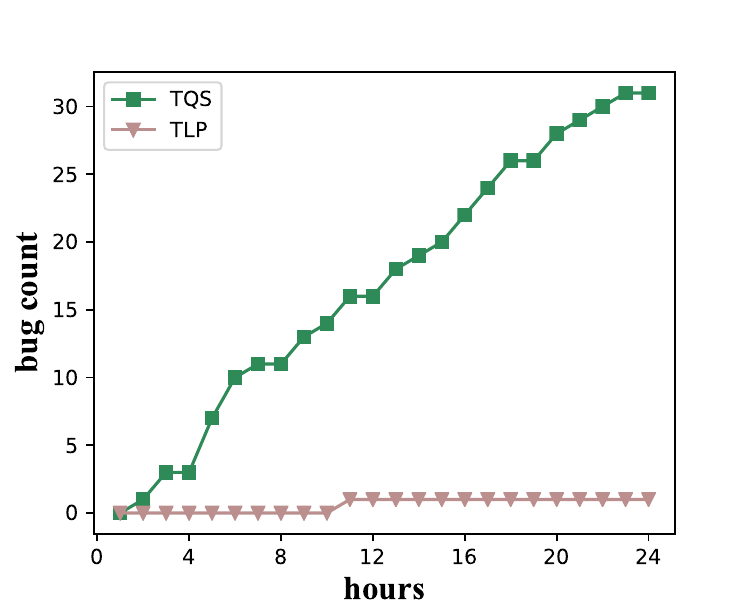}
      \end{minipage}
      \vspace{-0.1cm}
    }
    \subcaptionbox{X-DB efficiency}{
      \begin{minipage}[c]{.23\linewidth}
        \centering
        \includegraphics[scale=0.35]{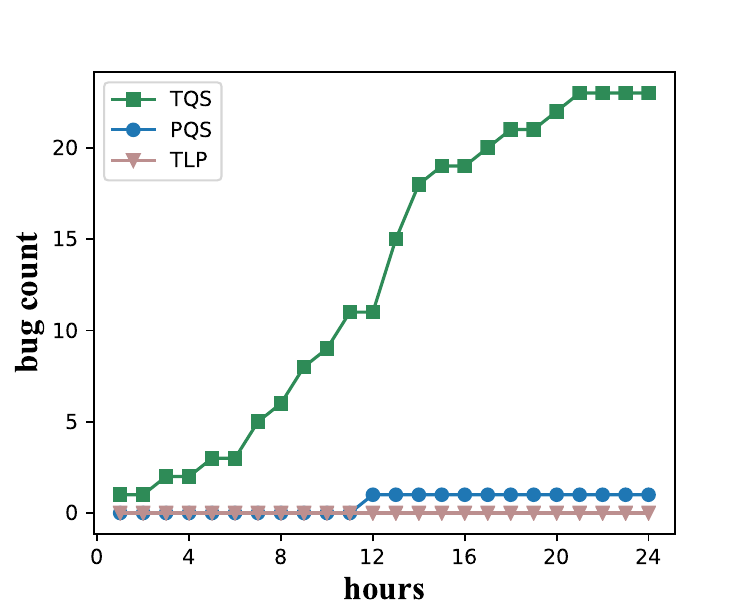}
      \end{minipage}
      \vspace{-0.1cm}
    }
    \vspace{-0.1cm} 
    \caption{Comparison with existing tools in the query graph diversity and the efficiency of detecting bugs.}
    \label{fig.compare}
    \vspace{-0.2cm} 
\end{figure*}

\subsubsection{Bugs in TiDB}
Merge join and join index are the main causes of bugs in TiDB.
Listing \ref{listing:tidb_varchar} 
shows an example query, where incorrect result set is obtained when hash join is converted to merge join (due to space constraints, the ground truth results and incorrect results are discarded here).

\begin{lstlisting}[language=SQL,
        caption={TiDB's merge join executed incorrect result set which misses -0.},
        label=listing:tidb_varchar]
CREATE TABLE t1 (
  id bigint(64) NOT NULL AUTO_INCREMENT,
  col1 varchar(511) DEFAULT NULL,
  PRIMARY KEY (id));

CREATE TABLE t2 (
  id bigint(64) NOT NULL AUTO_INCREMENT,
  col1 varchar(511) DEFAULT NULL,
  PRIMARY KEY (id));

CREATE TABLE t3 (
  id bigint(64) NOT NULL AUTO_INCREMENT,
  col1 varchar(511) DEFAULT NULL,
  PRIMARY KEY (id));

SELECT /*+ merge_join(t1, t2, t3)*/ t3.col1 FROM (t1 LEFT JOIN t2 ON t1.col1=t2.col1) LEFT JOIN t3 ON t2.col1=t3.col1;

SELECT /*+ hash_join(t1, t2, t3)*/ t3.col1 FROM (t1 LEFT JOIN t2 ON t1.col1=t2.col1) LEFT JOIN t3 ON t2.col1=t3.col1;
\end{lstlisting}

We find that when TiDB executes the above merge join, intermediate data 
are mistakenly materialized as null.
TiDB developers stated that it was because outer merge join cannot keep the prop of its inner child and the bug has been fixed.

\subsubsection{Bugs in X-DB}

In our X-DB, most bugs are produced from hash join and semi-join. 
Listing \ref{listing:X-DBleft} illustrates an example query which transforms left join to inner join.  X-DB fails to return a correct result set. 
We submit it to our developers and they located the cause of the bug: 
the inner join cannot distinguish null from 0. 
\begin{lstlisting}[language=SQL,
        caption={X-DB's incorrect inner join execution.},
        label=listing:X-DBleft]
CREATE TABLE t1 (
  id bigint(64) NOT NULL AUTO_INCREMENT,
  col1 int(16) NOT NULL,
  PRIMARY KEY (id, col1));

CREATE TABLE t2 (
  id bigint(64) NOT NULL AUTO_INCREMENT,
  col1 int(16) NOT NULL,
  PRIMARY KEY (id, col1));

CREATE TABLE t3 (
  id bigint(64) NOT NULL AUTO_INCREMENT,
  col1 varchar(511) DEFAULT NULL,
  PRIMARY KEY (id));

query:
SELECT t1.id FROM (t1 LEFT JOIN t2 ON t1.col1=t2.id) JOIN t3 ON t2.col1=t3.col1 where t1.col1 = 1;

ground truth:
empty set

transformed query:
SELECT /*+JOIN_ORDER(t3, t1, t2)*/ t1.id FROM (t1 LEFT JOIN t2 ON t1.col1=t2.id) JOIN t3 ON t2.col1=t3.col1 where t1.col1 = 1;

result:
+------+
| id   |
+------+
| NULL |
| NULL |
+------+
\end{lstlisting}

As another example, Listing \ref{listing:X-DBsemi} shows that a test case caused the semi-join to return the wrong result sets. 
When the query is executed by an inner semi hash join without materialization strategy, the X-DB returns the incorrect result. 
\begin{lstlisting}[language=SQL,
        caption={X-DB's incorrect hash join execution.},
        label=listing:X-DBsemi]
CREATE TABLE t0 (c0 float);
CREATE TABLE t1 (c0 float);

query: 
SELECT ALL t1.c0 FROM t1 RIGHT OUTER JOIN t0 ON t1.c0 = t0.c0 WHERE t1.c0 IN (SELECT t0.c0 FROM t0 WHERE (t1.c0 NOT IN (SELECT t1.c0 FROM t1))=(1) IN (t1.c0));

ground truth:
Empty set

transformed query:
SET optimizer_switch='materialization=off';
SELECT ALL t1.c0 FROM t1 RIGHT OUTER JOIN t0 ON t1.c0 = t0.c0 WHERE t1.c0 IN (SELECT t0.c0 FROM t0 WHERE (t1.c0 NOT IN (SELECT t1.c0 FROM t1 )) = (1) IN t1.c0);

result:
+-----------+
| c0        |
+-----------+
| 292269000 |
+-----------+
\end{lstlisting}

\subsection{Comparison with Existing Tools} \label{sec.comparison}
We compare TQS with three state-of-the-art DBMS testing approaches in different aspects to show its effectiveness and efficiency. Note that due to the compatibility problem, SQLancer implements different approaches on different databases (PQS and TLP on MySQL and X-DB; NoRec on MariaDB;  TLP on TiDB).  
Figure \ref{fig.compare} shows our comparison results. We adopt two metrics. The diversity of graphs shows the number of different isomorphic sets tested, and the 
bug type count represents the number of returned logic bug types.

\textbf{Query graph diversity.}  
Figure \ref{fig.compare}(a), (b), (c) and (d) show the query graph diversity (in thousands of isomorphic sets) of MySQL, MariaDB, TiDB and X-DB, respectively.
TQS significantly outperforms  baseline approaches on testing diversity.
This is because that (1) SQLancer approaches may generate random joins with empty results which are not usable for testing and (2) TQS adopts the KQE to avoid repeatedly testing the same query structure.

\textbf{Efficiency.}
Figure \ref{fig.compare}(e), (f), (g) and
(h) show the testing efficiency of MySQL, MariaDB, TiDB and X-DB, respectively. 
Not surprisingly, we obtain a similar result as the diversity experiment. More query structures are being tested, more bugs can be discovered. 
 
\begin{figure}[h]
  \centering
 \includegraphics[scale=0.35]{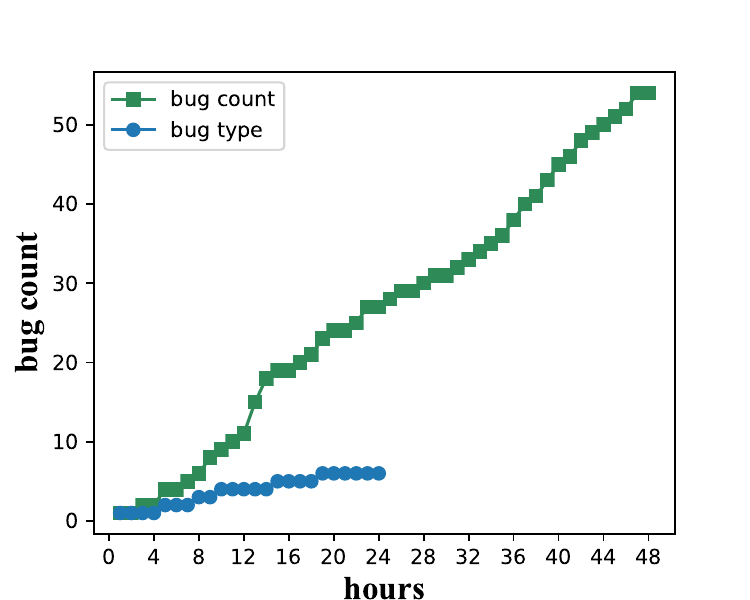}
  \caption{Bug types vs bug counts on MySQL.}
  \label{fig.runingtime}
\end{figure}

In fact, we run the TQS for 48 hours and continue to find logic bugs for all DBMSs. But there are too many bugs and we have not submitted them to the developer communities for verification.
So we only show bug types of the first 24 hours.
Figure \ref{fig.runingtime} illustrates that the number of bugs increases linearly with the testing time, while the number of bug types is not. This indicates that most bugs are caused by a small set of improperly implemented operators.

\begin{figure}[h]
  \centering
 \includegraphics[scale=0.35]{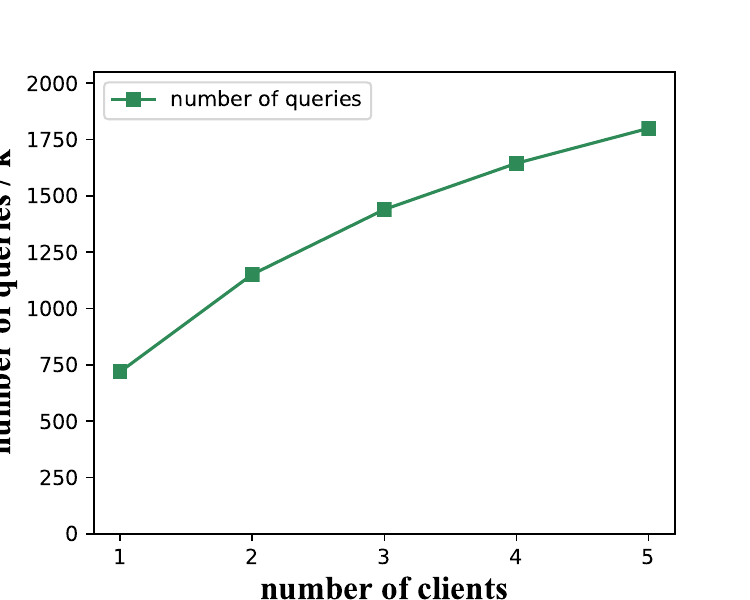}
  \caption{Effect of parallel search.}
  \label{fig.runingtime2}
\end{figure}

In Section \ref{sec:kqe}, we briefly explain the idea of using KQE to build a distributed computation framework to speed up the testing process. We show our results in Figure \ref{fig.runingtime2} by deploying the framework on the different number of clients. We run our experiments on MySQL for 24 hours and the results show that using parallel computation can effectively facilitate the testing process. 

\subsection{Ablation Studies} \label{sec.ablation}
In this section, we perform ablation experiments over some facets of TQS in order to better understand their roles.

\begin{table}
  \centering
    \caption{Ablation test over the effect of model composition.}
    \vspace{-0.2cm}
    \label{tab.ablation}
    \begin{tabular}{|c|c|c|c|}
    \hline
   DBMS &  Approach 
   & \makecell{Query Graph \\Diversity} &  \makecell{Bug Count} 
   \\
    \hline
      \multirow{4}{*}{\makecell{MySQL\\ 8.0.28}} & TQS & 460k  &  31 
      \\ \cline{2-4}
      & $TQS_{!Noise}$ & 460k & 14 \\ \cline{2-4}
      & $TQS_{!GT}$ & 460k & 21 \\ \cline{2-4} 
      & $TQS_{!KQE}$ & 228k & 16 \\
    \hline
      
    \multirow{4}{*}{\makecell{MariaDB\\ 10.8.2}} & TQS & 475k & {30}  \\ \cline{2-4}
      & $TQS_{!Noise}$ & 475k & 15 \\ \cline{2-4}
       & $TQS_{!GT}$ & 475k & 18  \\ \cline{2-4} 
      & $TQS_{!KQE}$ & 234k & 12  \\
    \hline
      
   \multirow{4}{*}{\makecell{TiDB\\ 5.4.0}} & TQS & 462k & 31 \\ \cline{2-4}
      & $TQS_{!Noise}$ & 462k & 20 \\ \cline{2-4}
    & $TQS_{!GT}$ & 462k & 22 \\ \cline{2-4} 
      & $TQS_{!KQE}$ & 231k & 18  \\
    \hline
      
    \multirow{4}{*}{\makecell{X-DB\\ 8.0.18}} & TQS & 465k & 23 \\ \cline{2-4}
      & $TQS_{!Noise}$ & 465k & 12  \\ \cline{2-4}
      & $TQS_{!GT}$ & 465k & 18 \\ \cline{2-4} 
      & $TQS_{!KQE}$ & 225k & 15 \\
    \hline
    \end{tabular}
     \vspace{-0.2cm}
    \end{table}

\textbf{Noise vs No-Noise.}
We artificially generate noise data during our testing.
As shown in Table \ref{tab.ablation}, the results indicate that the number of discovered bugs dramatically decreases, if we remove the noise-injection module (denoted as $TQS_{!Noise}$). 
This verifies that a large portion of logic bugs are generated by outliners or unexpected values. DBMS developers should be alerted of boundary testing.

\begin{lstlisting}[language=SQL,
        caption={X-DB's incorrect hash join execution.},
        label=listing:groundtruth]
CREATE TABLE t1 (
  id bigint(64) NOT NULL AUTO_INCREMENT,
  col1 int(16) DEFAULT NULL,
  col2 double DEFAULT NULL,
  PRIMARY KEY (id));

CREATE TABLE t2 (
  id bigint(64) NOT NULL AUTO_INCREMENT,
  col1 int(16) NOT NULL,
  col2 double DEFAULT NULL,
  col3 varchar(511) DEFAULT NULL,
  PRIMARY KEY (id, col1));

CREATE TABLE t3 (
  id bigint(64) NOT NULL AUTO_INCREMENT,
  col1 double DEFAULT NULL,
  col2 varchar(511) DEFAULT NULL,
  PRIMARY KEY (id));

SELECT t3.col2 FROM (t1 LEFT JOIN t2 ON t1.col1=t2.id) JOIN t3 ON t2.col1=tmp3.col3;
\end{lstlisting}

\textbf{GT vs No-GT.} 
 We first show the power of ground-truth verification. 
TQS without GT(ground-truth) is to judge the correctness of query results by comparing the results executed with different query plans (denoted as $TQS_{!GT}$), namely using the {\bf differential testing}.
We found that 7 bugs could not be detected using the differential testing on X-DB.
For example, Listing \ref{listing:groundtruth} shows a bug of hash join in X-DB 8.0.18. The query result remains the same for different plans. But the query result is different from the ground-truth of this query, 
indicating a logic bug here.
As shown in Table \ref{tab.ablation}, some bugs cannot be revealed by differential testing, while using ground-truth results, we can successfully identify them.

\textbf{KQE vs No-KQE.} 
KQE (knowledge-guided query space exploration) allows us to avoid the exploration of similar query structures.
In Table \ref{tab.ablation}, we observe that
TQS is superior to the $TQS_{!KQE}$ on the four databases, indicating the effectiveness of applying KQE to generate queries.
Note that because iterating all possible isomorphic sets for a graph is an NP-complete problem, there is no way to perform exhaustive testing. The intuition of KQE is to generate new queries as much as possible, not to iterate all isomorphic sets.

In summary, noise injection, ground-truth results and KQE modules are important for bug detection. They either improve the effectiveness of TQS, or speed up the testing process.

\section{Related Work}

\textbf{Differential Testing of DBMS.}
Differential testing is a widely adopted approach for detecting logic bugs in software systems.
It compares results of the same query from multiple versions of the system or uses different physical plans to discover possible bugs.
Differential testing has been shown to be effective in many areas \cite{brummayer2009fuzzing, DBLP:conf/nfm/CuoqMPPRYY12, DBLP:conf/pldi/LeAS14, DBLP:conf/kbse/KapusC17, DBLP:journals/dtj/McKeeman98, DBLP:conf/pldi/YangCER11}. 
It is first used in RAGS to find bugs of DBMSs \cite{DBLP:conf/vldb/Slutz98}. 
APOLLO also applies differential testing to find performance regression bugs by executing SQL queries on multiple versions of DBMSs \cite{DBLP:journals/pvldb/JungHAKK19}.
There were 10 previously unknown performance regression bugs found in SQLite and PostgreSQL. 
Although differential testing shows its effectiveness, our experiments show that some logic bugs must be revealed by using ground-truth results.  

\textbf{Generator-based Testing of DBMS.}
Various database data generators \cite{DBLP:conf/sigmod/BinnigKLO07,DBLP:conf/vldb/BrunoC05,DBLP:conf/vldb/HoukjaerTW06, DBLP:conf/kbse/KhalekELK08, DBLP:conf/sigmod/GraySEBW94} and query generators \cite{DBLP:journals/pvldb/JungHAKK19,DBLP:conf/sigmod/MishraKZ08,SQLSmith, DBLP:conf/sigmod/VartakRR10, DBLP:conf/vldb/BatiGHS07, DBLP:journals/tkde/BrunoCT06, DBLP:conf/vldb/Poess04} have been proposed to artificially create test cases, but test oracles, which should give feedback on the correctness of the system, have received less attention. 
Generation-based testings \cite{DBLP:journals/vldb/LoBKOH10, DBLP:conf/sigmod/MishraKZ08, DBLP:conf/kbse/KhalekK10, gu2012testing, DBLP:journals/sigmetrics/RehmannSHTBL16, DBLP:conf/sigmod/YanJJVL18, DBLP:conf/ccs/ZhongC0ZLW20} have been adopted for extensive testings on DBMSs  
for purposes such as bug-finding and benchmarking. 
SQLSmith is a widely-used generation-based DBMS tester \cite{SQLSmith}. It synthesizes a schema from initial databases and generates limited types of queries, whose target is at the code coverage. 
Squirrel focuses on generating queries  
to detect memory corruption bugs \cite{DBLP:conf/ccs/ZhongC0ZLW20}.
All above random generators are mostly applied to detect crashing bugs, while our focus is the logic bug.  

\textbf{Logic Bug Testing of DBMS.}
SQLancer \cite{SQLancer} is a current state-of-the-art tool in testing DBMS for logic bugs and is the most closely related work to ours.
SQLancer proposes three approaches to detect logic bugs.
PQS constructs queries to fetch a randomly selected tuple from a table \cite{rigger2020testing}.
TLP decomposes a query into three partitioning queries, each of which computes its result on a selected tuple \cite{rigger2020finding}. 
NoRec compares the results of randomly-generated optimized queries and rewritten queries that DBMS cannot optimize \cite{rigger2020detecting}. 
SQLancer targets at single table queries and 90.0\% of its bug reports include only one table. 
On the other hand, our TQS targets at detecting logic bugs of multi-table joins, which are more prone to bugs.

\textbf{Database Schema Normalization.}
The well-known database design framework for relational databases is centered around the notion of data redundancy \cite{DBLP:conf/sdb/Biskup95, DBLP:books/cs/Maier83}. The redundant data value occurrences originate from functional dependencies (FDs) \cite{DBLP:conf/sigmod/BiskupDB79}.
The data-driven normalization algorithms \cite{DBLP:journals/tods/DiederichM88, DBLP:conf/edbt/PapenbrockN17, DBLP:journals/pvldb/WeiL19} can remove FD-related redundancy effectively. Schema normalization has been well studied. 
In our approach, we adopt previous database normalization algorithm to spilt a wide table into multiple tables, so that the ground-truth results of join queries over those tables can be recovered from the wide table.

\textbf{Synthetic Graph Generation.} 
 In our approach, we adopt a random walk-based approach to generate valid join queries by enumerating sub-graphs from the schema graph. In fact, synthetic graph generation and the corresponding graph exploration approaches have been studied for years.  
For example, in support of the experimental study of graph data management system, a variety of synthetic graph tools such as SP2Bench~\cite{schmidt2009sp}, LDBC~\cite{erling2015ldbc}, LUBM~\cite{guo2005lubm}, BSBM~\cite{bizer2009berlin}, Grr~\cite{blum2011grr}, WatDiv~\cite{alucc2014diversified} and gMark~ \cite{DBLP:conf/icde/BaganBCFLA17} have been developed in the research community.  
In our scenario, iterating all possible sub-graphs is an NP-hard problem, and hence, we adopt a novel neural encoding-based approach to avoid generating similar sub-graphs, significantly reducing the overhead.

\section{Conclusion}
In this paper, we proposed a framework, TQS (\textbf{T}ransformed \textbf{Q}uery \textbf{S}ynthesis), for detecting logic bugs of the implementations of multi-table join queries in DBMS.
TQS employs two novel techniques, DSG (Data-guided Schema and query Generation) and KQE (Knowledge-guided Query space Exploration),
to generate effective SQL queries and their ground-truth results for testing.
We evaluated the TQS on four DBMSs: MySQL, MariaDB, TiDB and our cloud-native database X-DB. 
There are 115 
bugs discovered from tested DBMSs within $24$ hours.
Based on root cause analysis from the developer community, there are totally
7,5 
5 and 3 types of bugs in MySQL, MariaDB, TiDB and X-DB, respectively.
Compared to existing database debug tools, TQS
is more efficient and effective in detecting logic bugs generated by different join operators. It can be considered as an essential DBMS development tool.

\section*{Acknowledgment}
This work was supported by the Key Research Program of Zhejiang Province (Grant No. 2023C01037) and the Fundamental Research Funds for Alibaba Group through Alibaba Innovative Research (AIR) Program.

\bibliographystyle{ACM-Reference-Format}
\bibliography{main}

\end{document}